\documentclass[12pt]{iopart}

\pdfoutput=1

\usepackage[square,numbers,sort&compress]{natbib}
\usepackage[breaklinks=true,colorlinks=true,linkcolor=blue,urlcolor=blue,citecolor=blue]{hyperref}
\usepackage{latexsym}
\usepackage{graphics}
\usepackage{graphicx}
\usepackage{enumitem} 
\usepackage{epsfig}
\usepackage{mathcomp}
\usepackage{amsopn}
\usepackage{amsfonts}
\usepackage{amstext}
\usepackage{amssymb}
\usepackage{amstext}
\usepackage{subfigure}
\usepackage{braket}
\usepackage{iopams}  

\usepackage{iopams}
\usepackage[normalem]{ulem}

\usepackage[makeroom]{cancel}
\usepackage{epstopdf}
\usepackage{braket}

\usepackage{amsmath}
\usepackage[nointegrals]{wasysym}
\usepackage{physics}
\usepackage{xcolor}
\usepackage{comment}
\usepackage{natbib}
\usepackage{soul}
\usepackage{caption}

\def\url#1{}
\def\doi#1{}
\begin{document}

\title[Spin-J particles in magnetic fields]
{Quantum dynamics of spin-$J$ particles in static and rotating magnetic fields: Entanglement resonances and kinks}

\author{Nargis Sultana$^{*\,\dagger}$, 
Siddharth Seetharaman$^{\dagger}$ 
and Rejish Nath}

\address{Department of Physics, Indian Institute of Science Education and Research, Pune 411008, Maharashtra, India}

\begin{indented}
\item[] $^{*}$ Author to whom any correspondence should be addressed.
\item[] $^{\dagger}$ These authors contributed equally to this work.
\end{indented}
\ead{nargis.sultana@students.iiserpune.ac.in}
\begin{abstract}
We examine the quantum dynamics of both a single spin-$J$ particle and a pair of spin-$J$ particles in the presence of static and rotating magnetic fields, which can be important for qudit-based quantum technologies. Notably, we find resonant, periodic oscillations between two maximally stretched states, irrespective of the value of $J$. Additionally, we observe periodic transitions between sublevels with magnetic quantum numbers of opposite signs. The dynamics also exhibit a periodic transfer of the spin to the maximally stretched state, starting from the ground state of the initial Hamiltonian. For a pair of spins, we derive various resonance conditions and further analyze the entanglement generated by dipole-dipole interactions. In the case of two spin-1/2 particles, the entanglement dynamics reveal resonances and kinks in the maximum entanglement, and their criteria can be obtained from the energy spectrum. Strikingly, we show that the kink can be exploited to engineer the entanglement dynamics. Finally, we briefly discuss the regime of weak dipolar interactions, which are relevant for dipolar Bose-Einstein condensates.
\end{abstract}

%
%
%
%
%

\section{Introduction}
Understanding the behavior of spins in magnetic fields is a fundamental problem in classical and quantum physics \cite{levitt2008spin, PhysRevA.60.1824,PhysRevB.64.064410}, exhibiting fascinating features, ranging from Rabi oscillations in a spin-1/2 system \cite{Rabi_osc,cohen1977quantum} to exotic phase transitions \cite{suzuki2012quantum}. Various Hamiltonians describing spins in magnetic fields can be emulated in nuclear magnetic resonance (NMR) setups \cite{PhysRevA.71.012307, PhysRevB.75.094415,PhysRevLett.100.100501,RevModPhys.76.1037,PhysRevResearch.3.033035,10.1063/1.2838166,sciadv.adh2594}, in cold-atom/molecule systems \cite{cohen2012annual,moleculervw}, or in artificial quantum systems \cite{RevModPhys.86.153, PhysRevApplied.13.054059, PRXQuantum.2.017003}, such as quantum dots and superconducting circuits. The latest advancements in these systems are particularly exciting due to their potential applications in quantum technologies, enabled by the accessibility and controllability of quantum dynamics at the single-spin level. Consequently, there is a renewed interest in investigating the quantum dynamics of multi-level (spin) systems in external (magnetic) fields, especially when the fields are time-dependent, which can be utilized to manipulate the quantum states of these systems \cite{doi:10.1126/science.abd9547}.

Any two-level quantum system that acts as a qubit, with its energy levels coupled by external fields, can mimic the physics of a spin-1/2 in external magnetic fields. Similarly, systems with multiple energy levels form qudits \cite{quidt_review,yuxi25,vaarnmr} and may aid in understanding the properties of other large-spin complex systems \cite{PhysRevB.64.064410,Barra_1996,PhysRevLett.76.3830}. Laser-cooled atoms are particularly promising candidates for realizing spin-$J$ particles or qudits \cite{coherentcontrolhighdimensionalspace}. These cold atoms, which include alkali, alkaline-earth, and lanthanide types, exhibit rich hyperfine structures in their low-energy states. For instance, the ground state of bosonic Dysprosium (Dy) atoms has a total angular momentum of $J=8$ \cite{White_2025}. Additionally, a Dy atom also possesses a pair of quasi-degenerate states with opposite parity with $J=9$ and $J=10$ \cite{budker1993investigation,budker1994experimental,lepers2018ultracold}, with the latter being a metastable state. Such high-spin atoms possess large magnetic dipole moments, leading to strong dipole-dipole interactions (DDIs). These interactions can induce population mixing in atomic arrays \cite{PhysRevA.98.033815}, spin relaxation dynamics \cite{Hensler03, PhysRevLett.106.255303}, enhance the accuracy of atomic clocks \cite{PhysRevLett.127.013401}, and support highly entangled waveguide states in atomic arrays \cite{doi:10.1073/pnas.1911467116}. The role of DDIs has been explored in Bose-Einstein condensates \cite{PhysRevLett.124.073402,PhysRevA.106.063318} and is used to emulate exotic quantum spin models \cite{PhysRevA.110.023311}. The DDIs can be further tuned using rotating magnetic fields, especially their anisotropic character \cite{Tuning_DDI_Tilman_Pfau, Tuning_DDI_Lev}, which has far-reaching consequences in condensate physics \cite{PhysRevLett.95.200404,PhysRevA.76.013606,PhysRevLett.102.050401,PhysRevLett.100.240403,PhysRevLett.101.210402,PhysRevA.92.063601, PhysRevLett.122.050401, PhysRevA.101.043606, PhysRevA.100.023625,klaus22}. In general, multi-level systems exhibit various complex phenomena with potential applications. Some examples are, they can display complex absorption spectra \cite{Maguire_2006,SKOLNIK20091326},  non-classical mesoscopic-spin quantum states \cite{chalopin18}, affect the ac-Stark shifts \cite{Bender_2024}, suppress the EIT transparency \cite{10.1063/1.3269997,PhysRevA.83.053809}, suitable for exploring spinor physics \cite{KAWAGUCHI2012253}, facilitating the formation of long-lived collective dark states \cite{PhysRevLett.118.143602,PhysRevLett.123.223601,PhysRevA.101.043816,PhysRevX.12.011054}.


In light of recent developments in qudits and multi-level systems, in this paper, we study the dynamics of a single spin-$J$ particle and a pair of spins subjected to an external magnetic field, which is composed of a static field along the $z$-axis and an oscillating field in the $xy$-plane. These studies are particularly relevant for qudit based quantum applications. Moreover, a set of single-qudit and two-qudit gates can constitute fundamental building blocks for universal quantum computation \cite{quidt_review,ming14qcd,10.1063/1.1476391,PhysRevA.62.052309}. While the exact quantum dynamics of spins in static fields is well understood, the same in the presence of time-dependent fields—especially on multi-level atoms or spin-$J$ particles—remain largely unexamined. It was previously due to the lack of experimental setups capable of probing the dynamics at the level of a single or a couple of spins. However, as such capabilities have now emerged in experiments, we examine numerically and analytically the quantum dynamics of a single spin and a spin pair. The latter case is particularly interesting because correlations can be built between spins via DDIs. Although the exact analytical results for the quantum dynamics of a single spin can be obtained irrespective of the value of $J$, the introduction of a second spin, which leads to dipole-dipole interactions, complicates the problem significantly.

In the case of a single spin, we typically observe periodic dynamics. By mapping the spins onto a non-interacting gas of spin-1/2 particles, we obtain exact dynamical solutions \cite{landau1981quantum}. When the initial state is set to the lowest stretched state along the quantization axis, the spin undergoes periodic oscillations between the lowest and highest stretched states under a resonance condition, regardless of the value of $J$. Furthermore, if the initial state is one of the magnetic sublevels with a quantum number $m_j$, the dynamics exhibits periodic oscillations between the states $|m_j\rangle$ and $|-m_j\rangle$. Additionally, we show that it is possible to transfer the spin to a maximally stretched state, starting from the ground state of the initial Hamiltonian, which is generally a superposition of different $|m_j\rangle$ states.

Extending the analysis to a pair of spin-$J$ particles and incorporating their dipolar interactions, we examine the entanglement between the spins, both in the ground state and during the time evolution. As we discuss, there are non-trivial effects of the magnetic field on the correlation between the spins. Increasing the longitudinal field reduces the correlations between the spins in their ground state, while the dependence on the transverse field displays a non-monotonous behavior for small longitudinal field strengths. In the dynamics, starting from the lowest two-spin stretched state, the maximum entanglement as a function of field parameters exhibits resonant peaks. A careful analysis of the spin-1/2 case also reveals a kink, and we derive precise analytical conditions for its existence using the energy spectrum of the Hamiltonian in the rotating frame. Strikingly, we show that the entanglement kink can be used to engineer the entanglement dynamics in a controlled way in a pair of qubits. Furthermore, a stronger transverse field broadens the resonances, which can eventually lead to their overlap. Finally, we briefly discuss the regime of relatively weak dipolar interactions, which is particularly relevant for dipolar Bose-Einstein condensates.

The paper is organized as follows. Section~\ref{sm} discusses the Hamiltonian of a spin in a time-dependent magnetic field as well as the non-interacting model, where we break down the spin-$J$ particle into a gas of non-interacting spin-1/2 particles. The single spin dynamics for different initial states are covered in section~\ref{sdyn}. The properties of the ground state,  the population and entanglement dynamics, including the resonances, of two spins starting from a stretched state are discussed in section~\ref{spddi}. Finally, we briefly discuss the nature of dipolar interactions in the weakly interacting regime in section~\ref{wir}. The conclusion and outlook are provided in section~\ref{suou}. 

\section{Setup and models}
\label{sm}
\subsection{Hamiltonian of a spin in a magnetic field}
The Hamiltonian of a particle having total spin $J$ in a time-dependent magnetic field, $\mathbf{B} (t)$, is given by ($\hbar = 1$),
\begin{equation}\label{eq:Hamiltonian_gen}
    \hat{H} = - \hat{\vec{\mu}} \cdot \mathbf{B}(t) = g_J\mu_{B} \mathbf{B}(t) \cdot \hat{\mathbf{J}}
\end{equation}
where $\hat{\vec{\mu}} = - g_J \mu_B \mathbf{J}$ is the magnetic moment of the particle, $g_J$ is the g-factor and $\mu_B$ is the Bohr magneton. The model is equally valid for nuclear spins as in NMR setups. The magnetic field is given by $\mathbf{B}(t) = B_z \hat{z} + B_{\perp} \left[\cos \Omega t \ \hat{x} + \sin \Omega t \ \hat{y} \right]$, with its components $B_z$ and $B_\perp$ along and perpendicular to the $z$-axis. The axial component of the magnetic field is static, while the radial component rotates with a frequency, $\Omega$. By moving to a rotating frame defined by the unitary operator $\mathcal{U} = e^{i\Omega t \hat{J}_z}$, where $\hat{J}_z$ is the $z$-component of the angular momentum, we arrive at a time-independent Hamiltonian,
\begin{equation}\label{eq:HR}
    \hat{H}^\prime = \left( \omega_z - \Omega \right) \hat{J}_z + \omega_\perp \hat{J}_x, 
\end{equation}
 where $\omega_z = g_J \mu_B B_z$, $\omega_\perp = g_J \mu_B B_\perp$, and $\hat{J}_x$ is the $x$-component of the angular momentum. 

\subsection{Spin-$J$ particle as a non-interacting gas of spin-1/2 particles}
\label{nim}
We analytically study the dynamics of a spin-$J$ particle in a magnetic field governed by the Hamiltonian in equation~(\ref{eq:HR}) by representing it as a gas of $2J$ non-interacting spin-1/2 particles \cite{landau1981quantum}. The corresponding, many-particle spin-1/2 Hamiltonian is,
\begin{equation}
    \hat{H}^{(1/2)} = \frac{\omega_\perp}{2} \sum_{i=1}^{2J}\hat{\sigma}_x^i + \frac{(\omega_z-\Omega)}{2} \sum_{i=1}^{2J}\hat{\sigma}_z^i,
\end{equation}
where $\hat \sigma_x$ and $\hat \sigma_z$ are the Pauli spin-1/2 matrices. The total angular momentum operator is $\hat{\mathbf{J}}_{\text{tot}} = \sum\limits_{i=1}^{2J} \hat{\sigma}_i/2$, where $\hat\sigma=(\hat \sigma_x, \hat \sigma_y, \hat\sigma_z)$ and the eigenvalues corresponding to $\hat{J}_{\text{tot}}^2$ are $J_{\text{tot}}(J_{\text{tot}}+1)$ with $J_{\text{tot}}=J, J-1, ..., 1/2 \text{ or } 0$. The dynamics of $2J$ non-interacting spin-1/2 particles, governed by $\hat{H}^{(1/2)}$ in the subspace of $J_{\text{tot}}=J$, is equivalent to that of the single spin $J$ governed by the Hamiltonian in equation~(\ref{eq:HR}).
In this framework, the $|m_j = -J\rangle$ state of the original spin-$J$ particle corresponds to a state where all of the $2J$ spin-1/2 particles are in the down state, i.e., $|m_j = -J\rangle \leftrightarrow |\downarrow, \downarrow, \hdots, \downarrow \rangle$. On the other hand, the $|m_j = -J+n\rangle$ corresponds to the symmetric superposition of all the product states where $n$ of the $2J$ spin-1/2 particles are pointing up and the remaining ones are pointing down. 

Considering a general initial state of the $i$-th spin-1/2 particle, $\ket{\psi'_i(t=0)} = c_{i,\uparrow} \ket{\uparrow}_i + c_{i,\downarrow} \ket{\downarrow}_i$, the quantum state of that spin at a later time $t$ is given by
\begin{eqnarray}\label{eq:noninteracting-spin-1/2-rotating-frame}
    \ket{\psi_i'(t)} = 
    \left(c_{i,\uparrow} \cos \left( \dfrac{\omega^\prime t}{2}\right) - i \sin \left( \dfrac{\omega^\prime t}{2} \right) \left[ \dfrac{(\omega_z - \Omega)}{\omega^\prime} c_{i,\uparrow} + \dfrac{\omega_\perp}{\omega^\prime} c_{i,\downarrow} \right]\right) |\uparrow\rangle_i \nonumber \\ 
   +\left( c_{i,\downarrow} \cos \left( \dfrac{\omega^\prime t}{2}\right) + i \sin \left( \dfrac{\omega^\prime t}{2} \right) \left[ \dfrac{(\omega_z - \Omega)}{\omega^\prime} c_{i,\downarrow} - \dfrac{\omega_\perp}{\omega^\prime} c_{i,\uparrow} \right]\right)|\downarrow\rangle_i,
\end{eqnarray}
where $\omega'=\sqrt{(\omega_z - \Omega)^2 + \omega_\perp^2}$. Then, a tensor product of these single particle states yields the quantum state $|\Psi'(t)\rangle$ of the non-interacting gas of spin-1/2 particles. Finally, one can obtain the quantum state in the lab frame as $|\Psi(t)\rangle=e^{-i\Omega t \hat{J}_z} |\Psi'(t)\rangle$, where $e^{-i\Omega t \hat{J}_z}=\prod\limits_{i=1}^{2J} e^{-i \Omega t \hat{\sigma}_{z}^i/2}$, which imprints a dynamical relative phase factor between the spin-up and spin-down components in equation~(\ref{eq:noninteracting-spin-1/2-rotating-frame}).

\section{Single spin-dynamics: Analytical results}
\label{sdyn}
\subsection{Initial state: Lowest stretched state along the $z$-quantization axis} \label{sec:stretched_zquant}

\begin{figure}
    \centering
    \includegraphics[width=\linewidth]{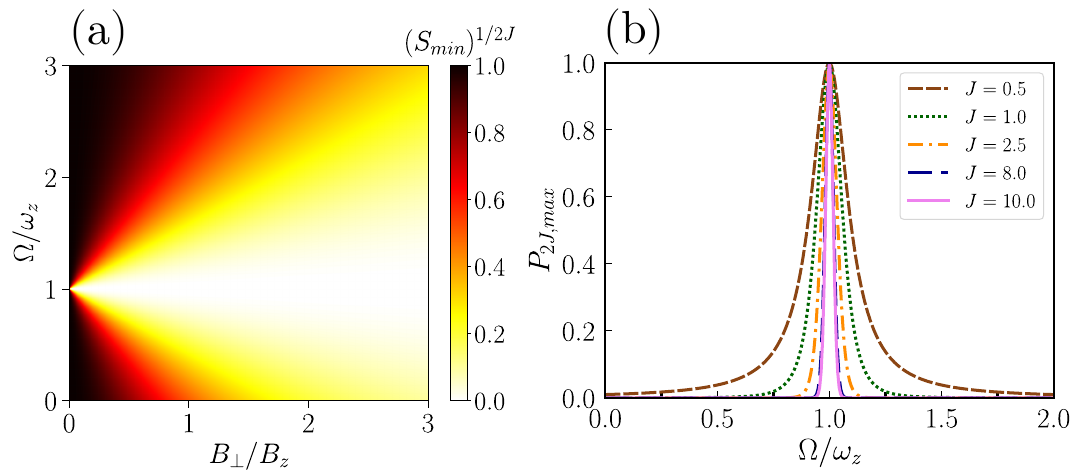}    
    \captionsetup{width=\linewidth} 
    \caption{(a) $(S_{\text{min}})^{1/2J}$ as a function of $\Omega/\omega_z$ and $B_\perp/B_z$. (b) shows maximum population attained in $|m_j = +J\rangle$ as a function of $\Omega/\omega_z$ for an initial state, $|\psi_0\rangle = |m_j = -J\rangle$ and $B_{\perp}/B_z=0.1$. As $J$ increases, it gets narrower and the tails decay rapidly.}
\label{f_smin}
\end{figure}

First, we discuss the dynamics for the initial state, $|\psi(t=0)\rangle = |m_j = -J\rangle$. Similar results can be obtained for $|m_j = J\rangle$. Labeling the $(2J+1)$ sublevels simply as $|n\rangle$ with $n=m_j+J$, where $n=0$ corresponds to $|m_j=-J\rangle$, the population in $\ket{n}$ as a function of time takes the form of binomial distributions and is given by [see \ref{app:population_levels_gen} for details of the calculation], 
 
\begin{equation}\label{eq:probabilities}
    P_n (t) = {}^{2J}C_n [p(t)]^n [q(t)]^{2J-n}
\end{equation}
where 
\begin{align}
    p(t) &= \dfrac{\omega_\perp^2}{\omega^{\prime 2}} \sin^2 \left( \dfrac{\omega^\prime t}{2} \right) \label{eq:p} 
\end{align}
and 
\begin{align}
    q(t) &= \left[ \cos^2 \left(\dfrac{\omega^\prime t}{2}\right) + \dfrac{(\omega_z - \Omega)^2}{\omega^{\prime 2}} \sin^2 \left(\dfrac{\omega^\prime t}{2}\right) \right]. \label{eq:q}
\end{align}
which results in periodic dynamics with time period, $T = 2\pi/\omega^\prime$. Taking $n=0$ in equation~(\ref{eq:probabilities}) gives us the survival probability of the initial state, which is simply $S(t) = q(t)^{2J}$, with the minimum value given by
\begin{equation}
    S_{\text{min}} = \left(1 - \dfrac{\omega_\perp^2}{(\omega_z - \Omega)^2 + \omega_\perp^2}\right)^{2J}.
\end{equation}
The appearance of the binomial distribution in the populations of the Zeeman sublevels can be closely related to that of spin-coherent states \cite{J_M_Radcliffe_1971}, as the time-evolution operator can be decomposed into a product of exponentials of $\hat{J}_\pm$ and $\hat{J}_z$ \cite{MARTINEZ_SU2_Decomposition}.

In figure~\ref{f_smin}(a), we show $(S_{\text{min}})^{1/2J}$ as a function of $B_\perp/B_z$ and $\Omega/\omega_z$, which captures features that are independent of the value of $J$. When the resonance condition $\Omega=\omega_z$ is satisfied, $S_{\text{min}}$ becomes zero for any non-zero value of $B_\perp$, indicating the spin oscillates between $|m_j=-J\rangle$, and $|m_j=J\rangle$, irrespective of how large the value of $J$ is, with a period $2\pi/\omega_\perp$. The frequency, $\omega_\perp$, depends on $J$ through $g_J$. Away from the resonance, as the rotation frequency increases, a larger $B_\perp$ is required to induce transitions. Hence, in the high-frequency limit, where $\Omega \gg \omega_z, \ \omega_\perp$ and low values of $B_\perp/B_z$, the value of $S_{\text{min}}$ almost remains unity, since the system does not get enough time to respond to the rotating field. In contrast, for large $B_\perp$, we observe the broadening of the resonant transition ($S_{\text{min}}=0$) along the $\Omega$-axis, about $\Omega=\omega_z$.

Further insights on the dynamics can be obtained by analyzing the maximum of $P_{n=2J}$, which is the probability of finding the spin in $|m_j=J\rangle$ state and is given by,
\begin{equation}\label{eq:max_prob_mj10}
    P_{2J, \text{max}} = \left[\dfrac{\omega_\perp^2}{\omega_\perp^2 + (\omega_z - \Omega)^2}\right]^{2J},
\end{equation}
which is a Lorentzian profile for $J=1/2$ [see figure~\ref{f_smin}(b)]. As $J$ increases, the central peak at $\Omega = \omega_z$ gets sharper and the tails damp out faster. We obtain the broadening or line width of this resonant transition as, $\Delta \Omega = (2^{1/2J} - 1)^{1/2} \omega_\perp$, which becomes increasingly smaller as $J$ increases. These resonant dynamics may be well suited for quantum sensing \cite{lkrt-lvng}.

Note that the magnetic moment of the atom can be obtained as $\vec{\mu} = - g_J \mu_B \langle \hat{\mathbf{J}} \rangle$, where $\langle \hat{\mathbf{J}} \rangle\equiv\langle \psi(t)|\hat{\mathbf{J}}|\psi(t)\rangle$, the expectation value of the angular momentum operator, and can be readily calculated [see \ref{app:dipole_moment_gen_stretched_state} for details]. The dynamics is then given by the precession of the dipole moment about the instantaneous direction of the effective magnetic field provided by the unit vector, $\hat e(t)=(\sin \theta_B \cos \Omega t, \sin \theta_B \sin \Omega t, \cos \theta_B)$, with $\theta_B = \tan^{-1} [\omega_\perp/(\omega_z - \Omega)]$. 

Similar features appear if we initialize the system in other Zeeman sublevels. At resonance, Rabi oscillations take place between $\ket{m_j = -J+n}$ and $\ket{m_j = J-n}$ at frequency $\omega^\prime$. The precession dynamics of the dipole moment is also identical to that of initial state $\ket{m_j = -J}$, albeit with a reduced magnitude for the dipole moment, $\mu = g_J \mu_B (J-n)$.

\subsection{Initial state: Ground state of the initial Hamiltonian} \label{sec:stretched_magnetic_field}

In this section, we analyze the dynamics starting from the ground state of the initial Hamiltonian [equation~(\ref{eq:Hamiltonian_gen}) or equation~(\ref{eq:HR}) with $\Omega=0$]. For small values of $B_\perp$, the ground state has its majority population in $|m_j=-J\rangle$, which would then give dynamics similar to that of the initial stretched state discussed earlier. As $B_\perp$ increases, mixing between different $m_j$ sublevels of the ground state becomes significant. In that case, we can expect a different and non-trivial population dynamics among the magnetic sublevels, $\{|m_j\rangle\}$. The initial magnetic field lies in the $xz$-plane, making an angle $\phi_0 = \tan^{-1} (B_{\perp}/B_z) = \tan^{-1} (\omega_\perp/\omega_z)$ with the $z$-axis.

\begin{figure}
    \centering
    \includegraphics[width=0.9\linewidth]{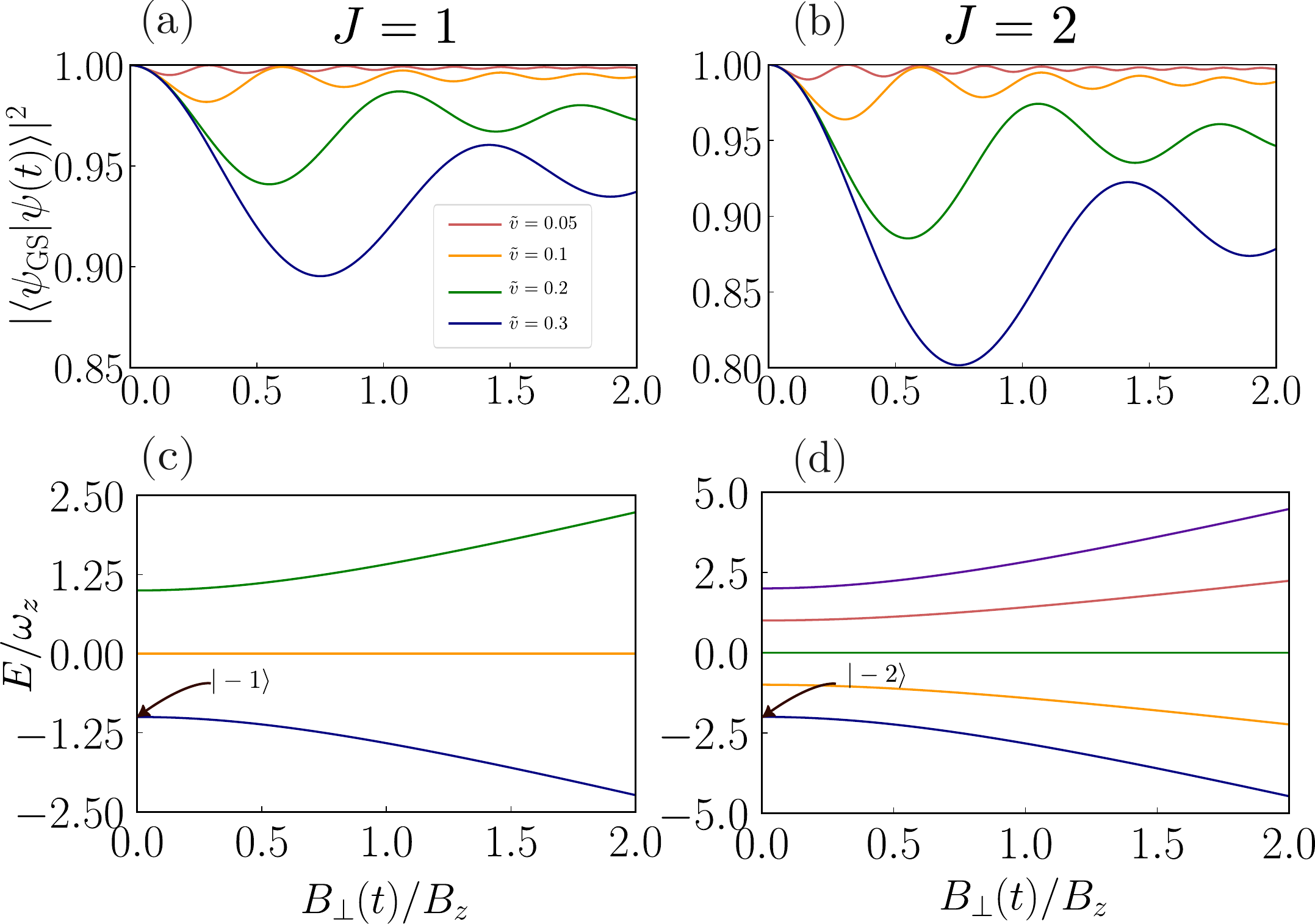}
    \caption{(a) and (b) The overlap between the time-evolved state and the instantaneous ground state under a linear quench of $B_\perp$ from $B_\perp=0$ to $B_\perp=2B_z$ (without any rotation, i.e. at $\Omega=0$) at different ramp rates for (a) $J=1$ and (b) $J=2$. The scaled quench rates, $\tilde v=v/\omega_z B_z$, for parts (a) and (b) are provided in the legend of (a). (c) and (d) show the corresponding energy eigenspectrum of $H(t=0)$ as a function of $B_\perp/B_z$ for $J=1$ and $2$, respectively.}
    \label{fig:adiabatic_ramp}
\end{figure}

{\em Adiabatic preparation.---} The ground state of the initial Hamiltonian can be prepared by adiabatically switching on the field along the $x$-axis while keeping $B_z$ fixed. In figure~\ref{fig:adiabatic_ramp}, we provide examples for $J=1$ and $J=2$. Particularly, we show the overlap of the time-evolved state $|\psi(t)\rangle$ with the instantaneous ground state $|\psi_{{\rm GS}}\rangle$ under the linear quench of $B_\perp$ with a rate, $v$, from $B_\perp=0$ to $B_\perp=2B_z$. The adiabatic condition can be obtained as $ Jv \ll \omega_z B_z$, which requires the product of $v$ and the coupling matrix elements to be much smaller than the energy gaps. Thus, the larger the value of $J$ and for fixed magnetic field strengths, the slower the quench must be to maintain the adiabatic dynamics, as also demonstrated in figure~\ref{fig:adiabatic_ramp}. 

\begin{figure}
    \centering
    \includegraphics[width=1\linewidth]{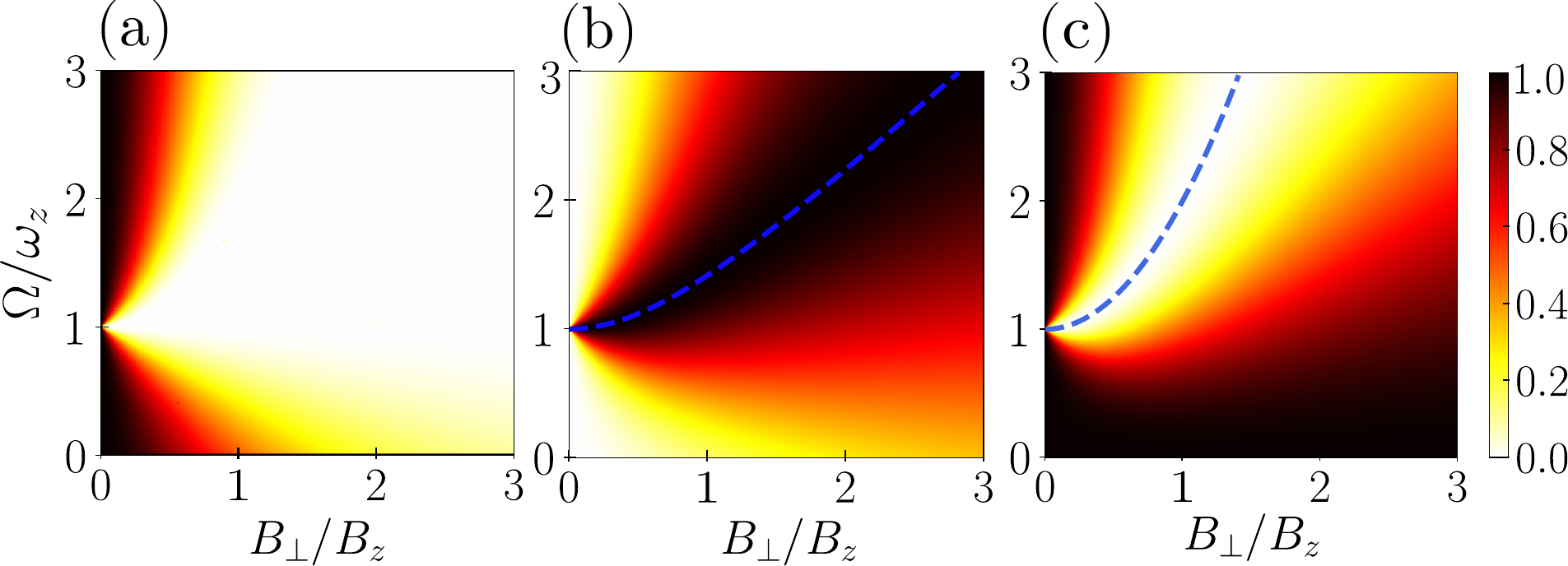}
    \caption{
   (a) $(S_{min})^{1/2J}$, (b) $(P_{2J,max})^{1/2J}$ and (c) $(P_{GS,min})^{1/2J}$ as a function of $\Omega/\omega_z$ and  $B_\perp /B_z$. The dashed line in (b) shows the criteria $\Omega/\omega_z=B/B_z$, where $B=\sqrt{B_\perp^2+B_z^2}$ and that in (c) corresponds to $\Omega/\omega_z=(B/B_z)^2$ where the overlap vanishes and the population is periodically transferred to the highest stretched state along the instantaneous magnetic field.}
    \label{sminp2j}
\end{figure}

Note that the initial state can be written as $\ket{\psi_0} = e^{-i\phi_0 \hat{J}_y} \ket{m_j = -J}$, where $\hat J_y$ is the $y$-component of the angular momentum operator. In terms of the states of $2J$ non-interacting spin-1/2 particles, the corresponding initial state is obtained by rotating each of the $2J$ spin-1/2 states (initially in down state) by $\phi_0$ about the $\hat{y}$-axis and is given by $\ket{\psi_0}_i = - \sin (\phi_0/2) \ket{\uparrow}_i + \cos (\phi_0/2) \ket{\downarrow}_i$. Then, the initial state of the whole system, $\ket{\Psi'(t=0)}$, is simply the tensor product of the individual ones. Then, using equation~(\ref{eq:noninteracting-spin-1/2-rotating-frame}), we obtain the survival probability of the initial state as (see \ref{app:survival_probability_gen})

\begin{equation}
    \begin{split}
        S (t) = \Bigg[ 1 &- \sin^2 \phi_0 \bigg( \cos^2 \left( \dfrac{\omega^\prime t}{2} \right) \sin^2 \left( \dfrac{\Omega t}{2} \right) \\
        &+ \sin^2 \left( \dfrac{\omega^\prime t}{2} \right) \left[\dfrac{\omega^2}{\omega^{\prime 2}} \sin^2 \left( \dfrac{\Omega t}{2} \right) + \dfrac{\Omega^2}{\omega^{\prime 2}} \cos^2 \left( \dfrac{\Omega t}{2} \right) \right] - \dfrac{\Omega}{2\omega^\prime} \sin (\omega^\prime t) \sin (\Omega t) \bigg) \Bigg]^{2J},
        \label{St3}
    \end{split}
\end{equation}
where $\omega = g_J \mu_B B$ with $B = \sqrt{B_\perp^2 + B_z^2}$ being the strength of the total magnetic field, $\mathbf{B}(t)$. Similarly, we also obtain the population in $\ket{m_j = J}$ or  $n=2J$ as (see \ref{app:population_levels_gen}),
\begin{equation}
        P_{2J} (t) = \left[ \sin^2 \dfrac{\phi_0}{2} + \dfrac{\Omega \omega \sin^2 \phi_0}{\omega^{\prime 2}} \sin^2 \left( \dfrac{\omega^\prime t}{2} \right) \right]^{2J},
        \label{p2j3}
\end{equation}
and the projection of $|\Psi(t)\rangle$ onto the instantaneous ground state $|\Psi_{GS}\rangle$ as (see \ref{app:survival_probability_gen}),
\begin{equation}\label{eq:P_GS}
    P_{GS} (t) = \left[ 1 - \dfrac{\Omega^2 \sin^2 \phi_0}{(\omega_z - \Omega)^2 + \omega_\perp^2} \sin^2 \left( \dfrac{\omega^\prime t}{2} \right) \right]^{2J}.
\end{equation}
Using equations~(\ref{St3})-(\ref{eq:P_GS}), we get critical insights into the spin-dynamics [see figure~\ref{sminp2j}]. 

\begin{figure}
    \centering
\includegraphics[width=0.7\linewidth]{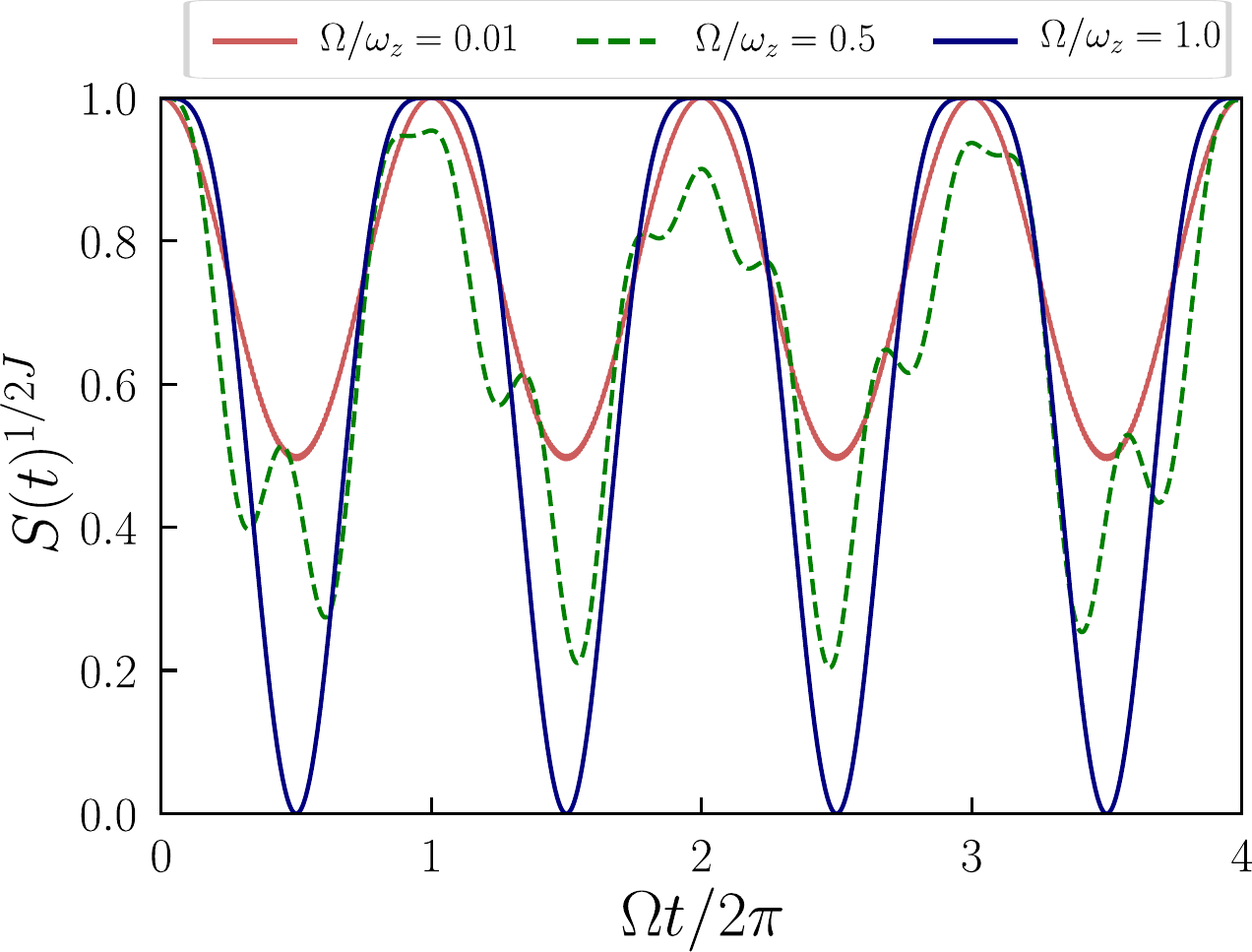}
    \caption{Variation of $S(t)^{1/2J}$ with time for $B_\perp/B_z=1$ for different rotation frequencies.}
    \label{fig:SPDYN}
\end{figure}

In figure~\ref{sminp2j}, we show  $(S_{min})^{1/2J}, \ (P_{GS, min})^{1/2J}$ and $(P_{2J,max})^{1/2J}$ as a function of $\Omega/\omega_z$ and $B_\perp/B_z$. For sufficiently large values of $B_\perp/B_z$ and $\Omega/\omega_z$, we observe that  $S_{min}\sim 0$ [see figure~\ref{sminp2j}(a)], which indicates that the spin periodically evolves into a state orthogonal to the initial state. This is confirmed by the dynamics of $S^{1/2J}(t)$ shown in figure~\ref{fig:SPDYN}, where we have taken $B_\perp/B_z=1$ and  observe that as $\Omega/\omega_z$ increases, $S^{1/2J}(t)$ periodically reaches a minimum of zero. Interestingly, as marked in figure~\ref{sminp2j}(b) by a dashed line, we see that when $\Omega/\omega_z=B/B_z$ (or equivalently, when $\Omega = \omega$), $P_{2J, max}\sim 1$. It indicates that the spin becomes fully polarized along the $z$-axis periodically, although the initial state is completely de-localized across the different $m_j$ sublevels. Since the fully polarized state $|m_j=J\rangle$ corresponds to all the spin-1/2 particles pointing upwards, the condition for $P_{2J, max}=1$ arises from ensuring that the contribution to $|\downarrow\rangle_i$ vanishes. Furthermore, as seen in figure~\ref{sminp2j}(b), the region where $P_{2J, max}\sim 1$ becomes broader with increasing $B_\perp/B_z$. These results suggest that it is indeed possible to coherently create an ensemble of atoms in the maximally stretched state, $|m_j=J\rangle$, along the $z$ axis, starting from a superposition of $m_j$ sublevels, using rotating magnetic fields, irrespective of how large $J$ is, considering the spin-spin interactions are negligible. 

In the regime of small rotation frequencies ($\Omega/\omega_z\ll 1$), the system adiabatically follows the instantaneous ground state, which makes $P_{GS, min}\sim 1$, as seen in figure~\ref{sminp2j}(c). In the adiabatic limit, we have $[S(t)]^{1/2J} \sim 1 - \sin^2 \phi_0 \sin^2 (\Omega t/2)$, which depends on the magnetic fields through $\phi_0$. In particular, the minimum of survival probability decreases with increasing $B_\perp$ while keeping $B_z$ constant, and approaches zero in the limit $B_\perp/B_z\to\infty$. 
 
As shown in figure~\ref{sminp2j}(c), the minimum population in the instantaneous ground state becomes zero when $\Omega/\omega_z= (B^2/B_z^2)$ (marked by a dashed line). The population dynamics reveals that, under this condition, the spin gets periodically transitioned into the highest stretched state along the direction of the magnetic field in the rotating frame, or the direction of the magnetic field at that instant of time in the lab frame. In the rotating frame, the latter corresponds to the direction of the initial magnetic field, $\mathbf{B}(t=0)$, which is set by the angle, $\phi_0 = \tan^{-1}(\omega_\perp/\omega_z)$ with respect to the $z$-axis and lying in the $xz$ plane. Hence, this dynamics can be understood by a unitary transformation, where we change the basis from the eigenstates of $\hat J_z$ to that of an axis (say $z'$) along the direction of $\mathbf{B}(t=0)$. Under this transformation, the Hamiltonian in equation (\ref{eq:HR}) becomes,
\begin{equation}
\label{Hrot}
    \hat{H} = \left[ \left(\omega_z - \Omega \right) \cos \phi_0 + \omega_z \dfrac{\sin^2 \phi_0}{\cos \phi_0} \right] \hat{J}_{z^\prime} +  \Omega  \sin \phi_0 \hat{J}_{x^\prime},
\end{equation}
where $\hat J_{z'}$ and $\hat J_{x'}$ are the components of the transformed angular momentum operator. Now, it is easier to see that when the first term in equation~(\ref{Hrot}) vanishes (equivalently $\Omega/\omega_z= (B^2/B_z^2)$), the system oscillates between the stretched states $\ket{m_j^\prime = -J}$ and $\ket{m_{j}^\prime = +J}$.

As with the initial state $\ket{m_j = -J}$, the expectation value of the magnetic moment precesses about the instantaneous direction of the resultant magnetic field. In the adiabatic limit when $\Omega/\omega \rightarrow 0$, $\theta_B \rightarrow \phi_0$ and the magnetic moment is oriented along the instantaneous direction of the magnetic field and precesses about the $z$-axis, as expected. 

\section{Two spins}
\label{spddi}

In this section, we discuss the ground state properties and quantum dynamics of a pair of spin-$J$ particles subjected to static and rotating magnetic fields. Assuming the spins are frozen in space, hence, neglecting the motional dynamics, the Hamiltonian of the system can be written as, 
\begin{equation}
    \hat{H} = - \sum\limits_{i=1}^2 \hat{\vec{\mu}}_i \cdot \mathbf{B} (t) + \hat{V}_{dd} (\vec{r})
    \label{h2s}
\end{equation}
where $\vec{\mu}_i = - g_J \mu_B \mathbf{J}_i$ is the magnetic moment of each spin. $V_{dd} (\vec{r})$ is the dipolar potential between them separated by a radial vector $\vec r=r\hat r$, which takes the form \cite{Hensler03},
\begin{equation}\label{eq:V_DDI}
    \hat{V}_{dd} (\vec{r}) = \dfrac{\mu_0}{4\pi}\left[\dfrac{\hat{\vec{\mu}}_1 \cdot \hat{\vec{\mu}}_2 - 3 (\hat{\vec{\mu}}_1 \cdot \hat{r}) (\hat{\vec{\mu}}_2 \cdot \hat{r})}{r^3}\right],
\end{equation}
where $\mu_0$ is the vacuum permeability. The distance $r$ can be varied to control the strength of DDIs. We diagonalize the Hamiltonian in equation~(\ref{h2s}) in a space spanned by the product states $\ket{m_{j_1}, m_{j_2}}$, which are the eigenstates of the $z$-component of the the total angular momentum and has a dimensionality of $(2J+1)^2$. In the following, we assume that the atoms are placed along the $z$-axis ($\hat r=\hat z$), and the magnetic dipolar potential becomes
 \begin{equation}
    V_{dd}=g_d\left[\mathbf{\hat {J}}_{1} \cdot \mathbf{\hat {J}}_{2}-3\hat {J}_{1z}\hat {J}_{2z}\right],
    \label{vdz}
\end{equation}
where $g_d=\mu_0(g_J\mu_B)^2/(4\pi r^3)$ is the dipolar interaction strength between the spins, $\hat J_{1\alpha}$ and $\hat J_{2\alpha}$ are the spin operators of the first and second dipoles. Note that the strength of DDIs is proportional to $J^2$, whereas the Zeeman terms depend linearly on $J$. Below, we numerically solve the Schr\"odinger equation $i\partial_t \ket{\psi (t)} = \hat{H} (t) \ket{\psi (t)}$ to study the dynamics. Note that in the presence of interactions, it is no longer convenient to map each of the spin-$J$ particles to an ensemble of $2J$ spin-1/2 particles, which would then require all-to-all interactions between the spins in one ensemble with the spins in the other.

More insights into the dynamics can be gained in the rotating frame defined by $\hat{U}=\exp[i \hat J_z\Omega t]$, where $\hat J_z=\hat{J}_{1z}+\hat{J}_{2z}$, and the Hamiltonian becomes,
\begin{align}
    \hat H_\text{rot}/g_d & = \left(\beta_z-\dfrac{\Omega}{g_d} \right)\hat J_z + \beta_\perp \hat J_x + ( \mathbf{\hat{J}}_1 \cdot \mathbf{\hat{J}}_2 - 3\hat J_{1z}\hat J_{2z}),
    \label{hrot}
\end{align}
where $\hat J_x=\hat{J}_{1x}+\hat{J}_{2x}$ and the dimensionless parameters, $\beta_z=g_J \mu_BB_z/g_d$ and $\beta_\perp=g_J \mu_BB_\perp/g_d$, quantify the relative strengths of the Zeeman terms with respect to the dipolar interaction strength. The dipolar potential leads to exchange and Ising terms, with the former being essential for realizing two-qubit entangling gates \cite{divi00,vel15,kand19,wat18,doi:10.1126/science.aao5965}. If $\hat{r} \neq \hat{z}$, the Hamiltonian becomes time-dependent even in the rotating frame since the unitary transformation to the rotating frame does not commute with $(\hat{\mathbf{J}}_1 \cdot \hat{r})(\hat{\mathbf{J}}_2 \cdot \hat{r})$ in this case.

\subsection{Ground state properties ($\Omega=0$)}
\label{gsp}

\begin{figure}
    \centering
    \includegraphics[width=1\linewidth]{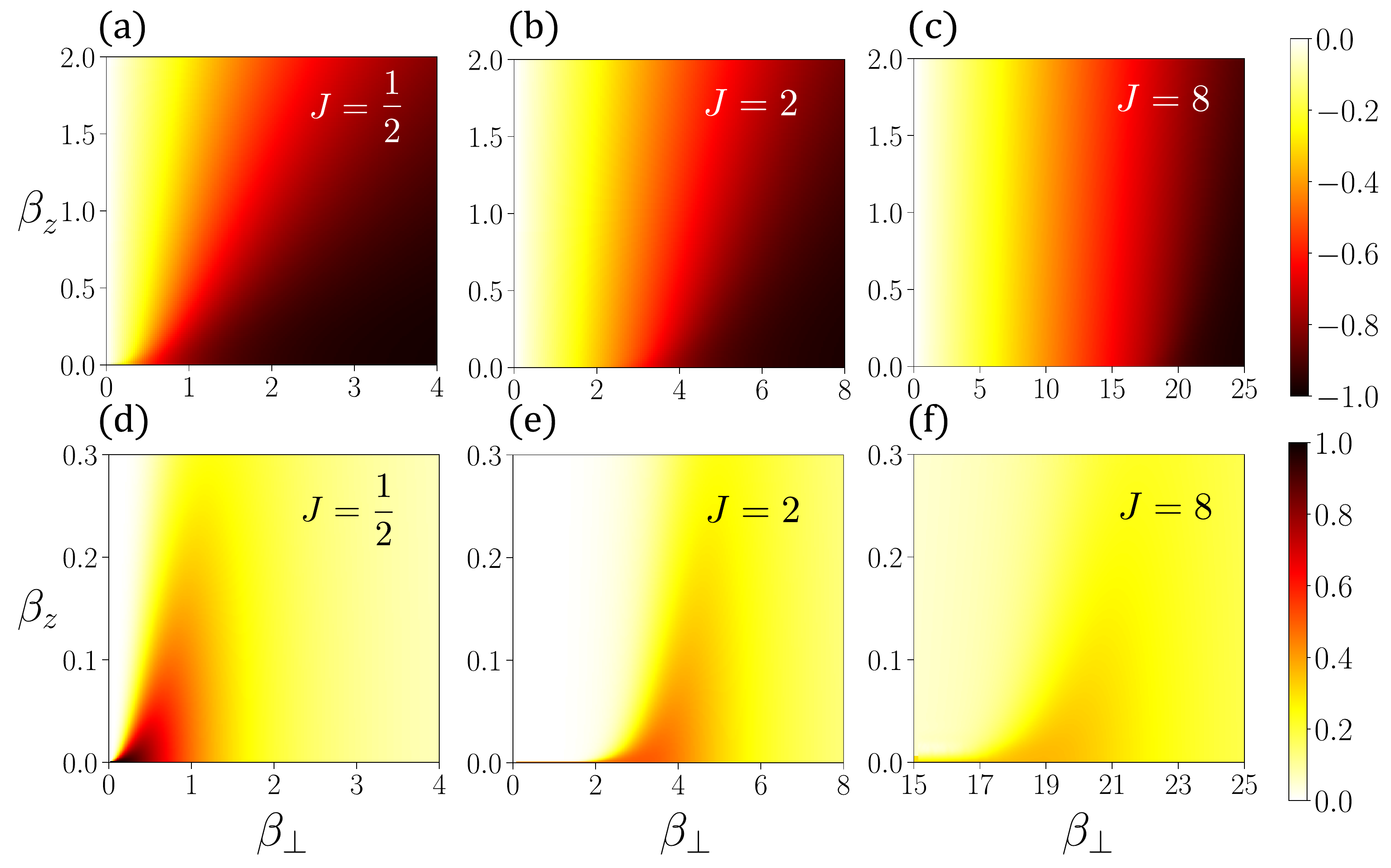}
    \caption{Ground state properties of a spin pair. (a)-(c) show $\langle J_x \rangle/2J$ of the ground state of the Hamiltonian in equation~(\ref{h2s}) at $t=0$ as a function of field strengths for different $J$. (d)-(f) show the corresponding entanglement entropy of one spin.}
    \label{fig:4}
\end{figure}

\begin{figure}
    \centering
    \includegraphics[width=0.7\linewidth]{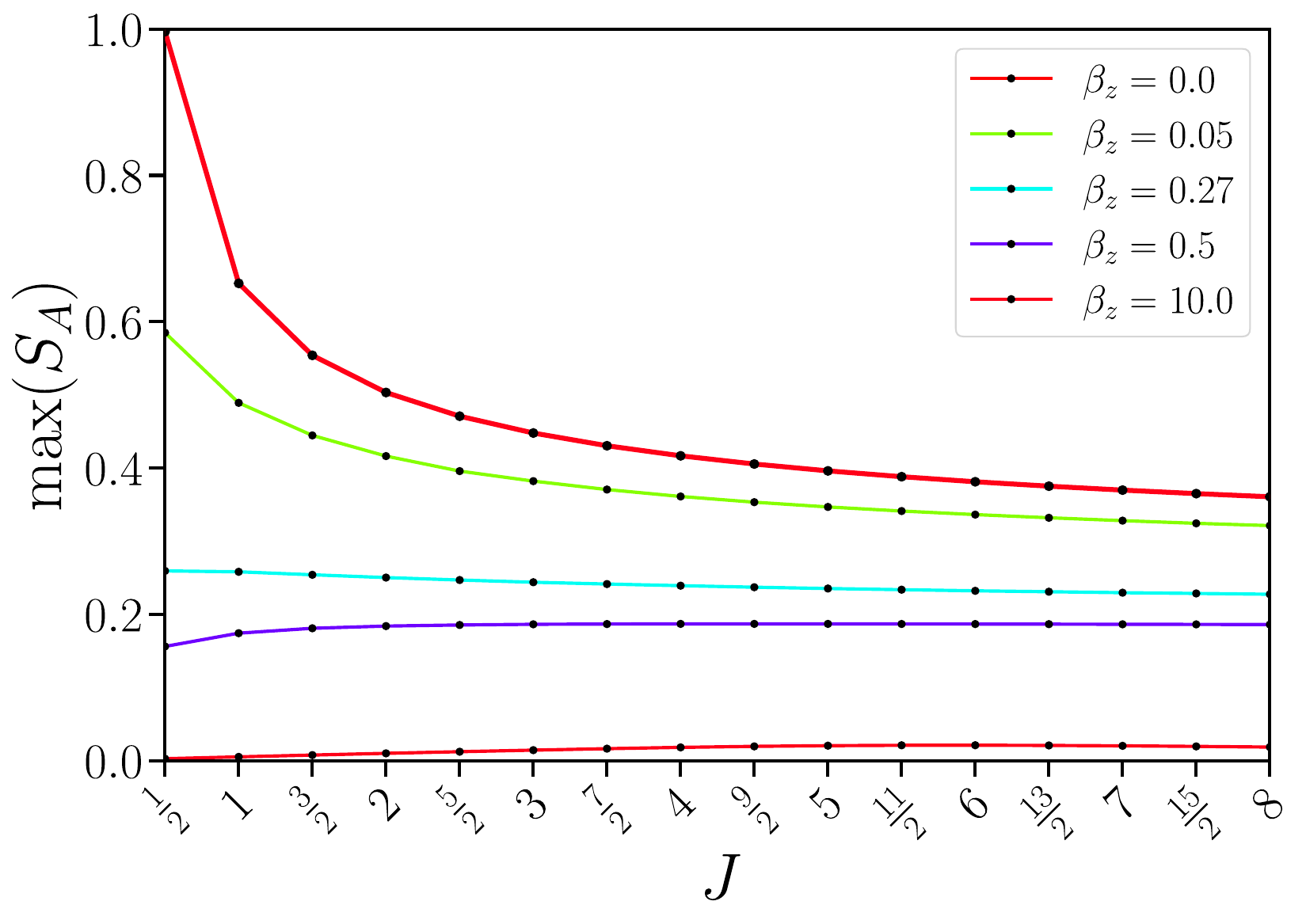}
    \caption{The maximum entanglement entropy of the ground state obtained for different $J$ by varying $\beta_\perp$ at fixed values of $\beta_z$.}
    \label{fig:SAmaximum}
\end{figure}

Before indulging in the quantum dynamics of two spin-$J$ particles, we examine the ground states for $\Omega=0$. When the dipolar interaction dominates over the Zeeman terms, the term $-3\hat J_{1z}\hat J_{2z}$ in equation~(\ref{vdz}) is more significant than the other two terms. It favors the ground states to be the stretched states, $\ket{\pm J,\pm J}$, and this double degeneracy is lifted by the Zeeman terms. Specifically, a positive $B_z$ favors $\ket{-J,-J}$ to be the ground state over $\ket{+J,+J}$ and vice versa. At the same time, the $B_\perp$ field admixes $\ket{\pm J,\pm J}$ with $\ket{\pm J\mp 1,\pm J}$ and $\ket{\pm J,\pm J\mp 1}$, and so on. 

To characterize the ground state properties, we look at $\langle \hat J_x\rangle$ and the entanglement entropy, $S_A = - \Tr(\hat\rho_A \ln_{2J+1}\hat\rho_A)$ [see figure~\ref{fig:4}], where $\hat\rho_A$ is the reduced density matrix of the subsystem, comprising of one of the spins. We use a logarithmic function with a base of $2J+1$ for the spin-$J$ particles, such that $S_A$ can only reach a maximum value of one, regardless of the value of $J$. As shown in figures~\ref{fig:4}(a)-\ref{fig:4}(c), irrespective of the value of $J$, when $\beta_\perp$ is sufficiently large and $\beta_z$ is small, both dipoles polarize maximally along the negative $x$ axis as expected, resulting in $\langle \hat J_x\rangle\sim -2J$ and the spins are unentangled or weakly entangled [see figures~\ref{fig:4}(d)-\ref{fig:4}(f)]. Conversely, when $\beta_z$ is large and $\beta_\perp$ is sufficiently small, the dipoles align along the negative $z$-axis and are again unentangled. In that case, $\langle \hat J_x\rangle=0$ and $\langle \hat J_z\rangle=-2J$. When both $\beta_z$ and  $\beta_\perp$ are large --- i.e. when the Zeeman terms dominate the DDIs --- there is minimal correlation between the two spins, and they align along the resultant magnetic field. As seen in figures~\ref{fig:4}(d)-\ref{fig:4}(f), for a given $J$, the maximum value of $S_A$ is attained when $\beta_z=0$ with a moderate value of $\beta_\perp$. The maximum value of one is attained only for $J=1/2$, in which the ground state is the Bell state, $\dfrac{1}{\sqrt{2}}\Big(\ket{\downarrow \downarrow}+\ket{\uparrow \uparrow}\Big)$, where $|\downarrow\rangle=|m_j=-1/2\rangle$ and $|\uparrow\rangle=|m_j=1/2\rangle$. In general, $S_A$ decreases as $\beta_z$ increases, because the Zeeman shifts from $\beta_z$ prevent the states from getting maximally mixed by the transverse magnetic field and dipolar exchange terms. In addition, $S_A$ exhibits a non-monotonic behavior as a function of $\beta_\perp$ for small values of $\beta_z$, which can be understood as follows: for small $\beta_z$, the DDI together with the transverse field $\beta_\perp$ builds up the correlations until the latter overwhelms the former. When $\beta_\perp$ is sufficiently large, the effect of DDI gets weaker, reducing the correlation between the dipoles.

\begin{figure}
    \centering
    \includegraphics[width=1\linewidth]{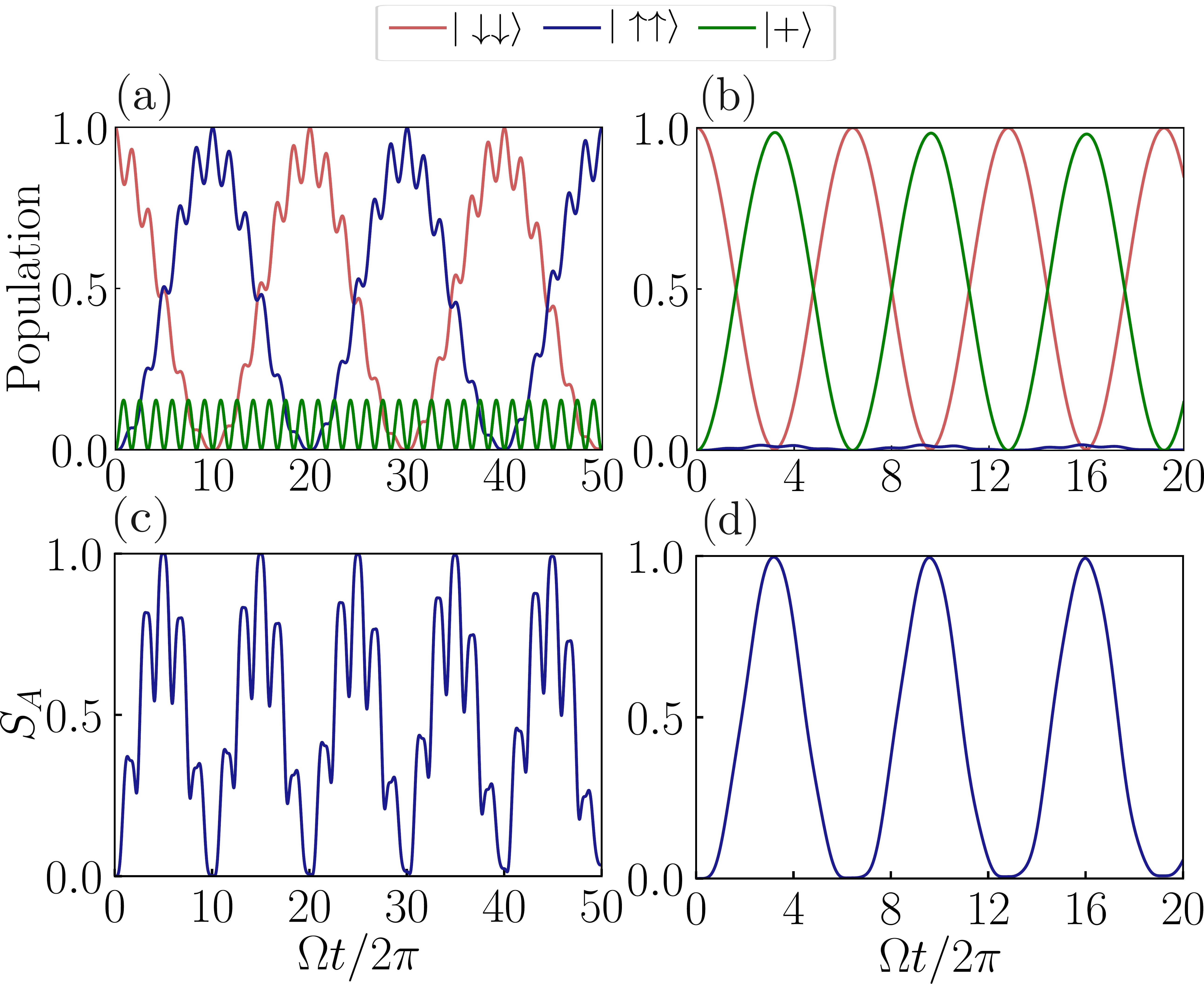}
    \caption{Resonant dynamics of two $J=1/2$ spins at $\beta_z = 3$ and $\beta_\perp = 0.5$. (a) and (b) show population dynamics corresponding to the two resonances discussed in the main text. The corresponding dynamics of entanglement entropy $S_A$ are shown in (c) and (d). (a) and (c) are for $\Omega=3 g_d$ (first resonance), and (b) and (d) are for $\Omega=4.5 g_d$ (second resonance).}
    \label{fig:2spd}
\end{figure}

As evident from the results shown in figure~\ref{fig:4}, larger the value of $J$, the stronger the DDIs ($\propto J^2$), which demands a larger $\beta_\perp$ to mix the nearby sublevels and generate significant entanglement between the two spins. The dependence of the maximum of $S_A$ on the spin $J$ is shown in figure~\ref{fig:SAmaximum}, which is obtained by scanning $\beta_\perp$ for different $\beta_z$.  As discussed above, the entanglement between the spins diminishes as the strength of the longitudinal field, $\beta_z$, increases. For small values of $\beta_z$, $(S_A)_{\rm max}$ decreases as $J$ increases, and around $\beta_z\sim 0.27$ it becomes nearly independent of $J$, and is $\sim 0.24$. For $\beta_z> 0.27$, $(S_A)_{\rm max}$ initially shows an increment but then remains almost independent of $J$. Crucially, the regions in figures~\ref{fig:4}(d)-\ref{fig:4}(f) where $S_A$ is significant provide us the range of $\beta_z$ and $\beta_\perp$ for which the DDI is relevant, as far as the ground state is concerned.

\subsection{Spin dynamics: Entanglement resonances and kinks}

Now, we analyze the quantum dynamics of the two spin-$J$ particles in combined static and rotating fields, starting from the initial state $|-J, -J\rangle$. It is apparent from the Hamiltonian that a finite $\beta_\perp$ and the dipolar exchange term can drive the system into the dynamics. 

\subsubsection{$J=1/2$.}
\label{dyj1-2}

\begin{figure}
    \centering
    \includegraphics[width=1\linewidth]{fig_entanglement_res_spin_0.5_copy.pdf}
    \caption{Maximum of entanglement entropy as a function of $\Omega$ for different $\beta_\perp$ and $\beta_z$.  (a)-(d) show the results for $\beta_\perp=$ 0.1, 0.5, 1, and 2, respectively. The maximum is obtained from analyzing the dynamics over a period of $T=300(2\pi/g_d\beta_\perp)$.}
    \label{fig:smax}
\end{figure}
First, we discuss the population and entanglement dynamics for $J=1/2$. A finite $\beta_\perp$ leads to resonant transitions between $\ket{\downarrow\downarrow}$  and $\ket{\uparrow\uparrow}$ when $\Omega/g_d = \beta_z$ [first resonance, see figure~\ref{fig:2spd}(a)], and between $\ket{\downarrow\downarrow}$ and $\ket{+}=(\ket{\uparrow\downarrow}+\ket{\downarrow\uparrow})/\sqrt{2}$ when $\Omega/g_d = \beta_z + 3/2$ [second resonance, see figure~\ref{fig:2spd}(b)]. The corresponding entanglement dynamics is shown in figures~\ref{fig:2spd}(c) and \ref{fig:2spd}(d), respectively. For the first resonance, the two spins become maximally entangled as they transition into the state $(\ket{\uparrow\uparrow}+\ket{\downarrow\downarrow})/\sqrt{2}$, while in the second resonance, it happens when they transition into $\ket{+}$. Figure~\ref{fig:smax} shows the maximum entanglement attained during the dynamics as a function of $\Omega/g_d$ for different $\beta_\perp$ and $\beta_z$. For small values of $\beta_\perp$, $(S_A)_{\rm max}$ exhibits two peaks as a function of $\Omega/g_d$, corresponding to the two resonances discussed above [see figure~\ref{fig:smax}(a)], which we refer to as entanglement resonances. The first resonance is sharper than the second since the former involves a second-order process with two spin flips. A finite $\beta_z$ shifts the resonances to larger values of $\Omega$, as shown by dashed lines. As $\beta_\perp$ increases [see figures~\ref{fig:smax}(b)-\ref{fig:smax}(d)], both resonances become broader, eventually merge and become indistinguishable at sufficiently large $\beta_\perp$.

\begin{figure}
    \centering
    \includegraphics[width=1\linewidth]{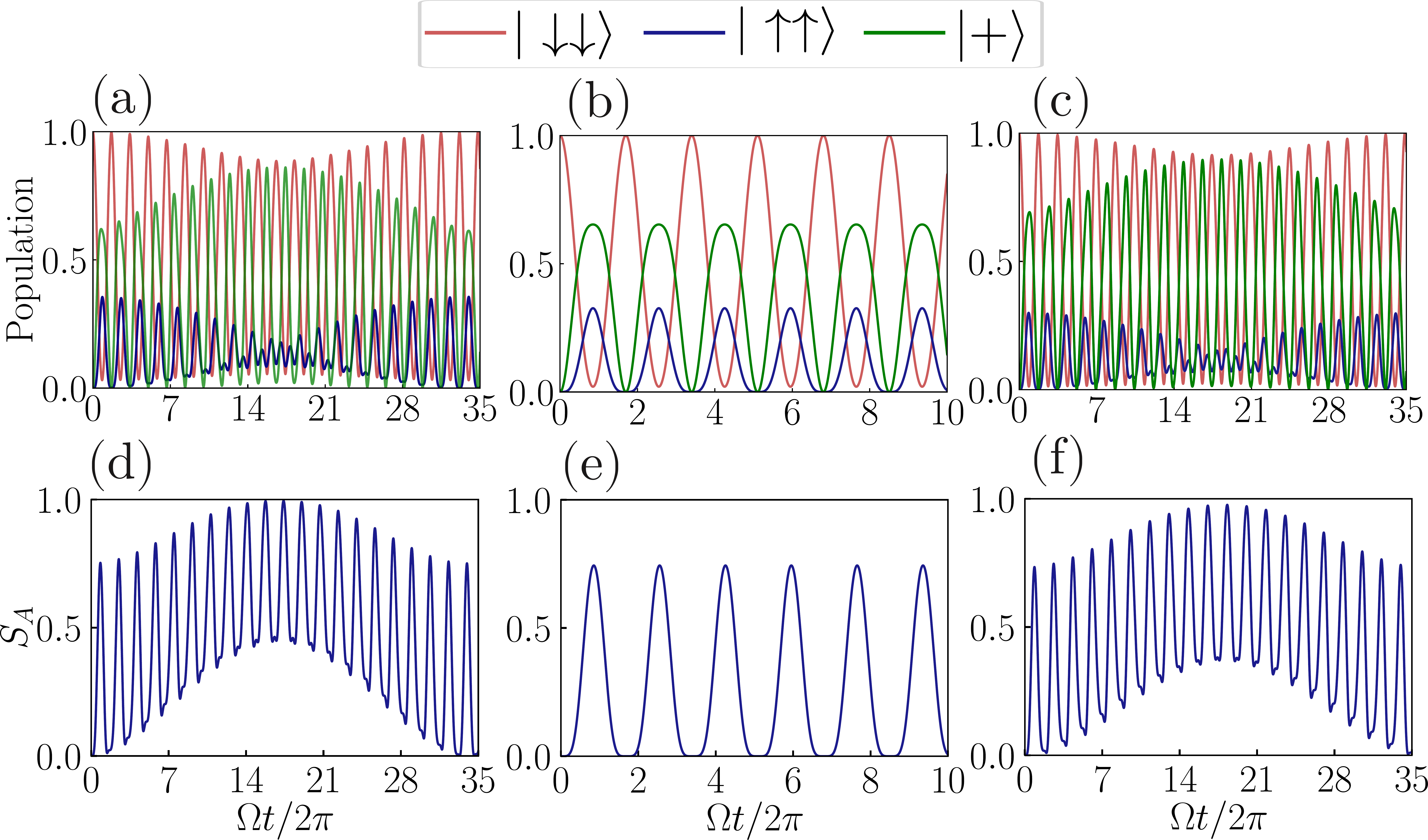}
    \caption{Population and entanglement dynamics for $\beta_z=3,\beta_\perp=2$ for different rotation frequencies, $\Omega/g_d=4.4,4.5$ and $4.6$ respectively. The dynamics shown in the middle column are characterized by a single frequency, provided by the energy difference, $E_3-E_1=E_4-E_3$, at the kink.}
    \label{fig:Omegadyn}
\end{figure}
\begin{figure}
    \centering
    \includegraphics[width=1\linewidth]{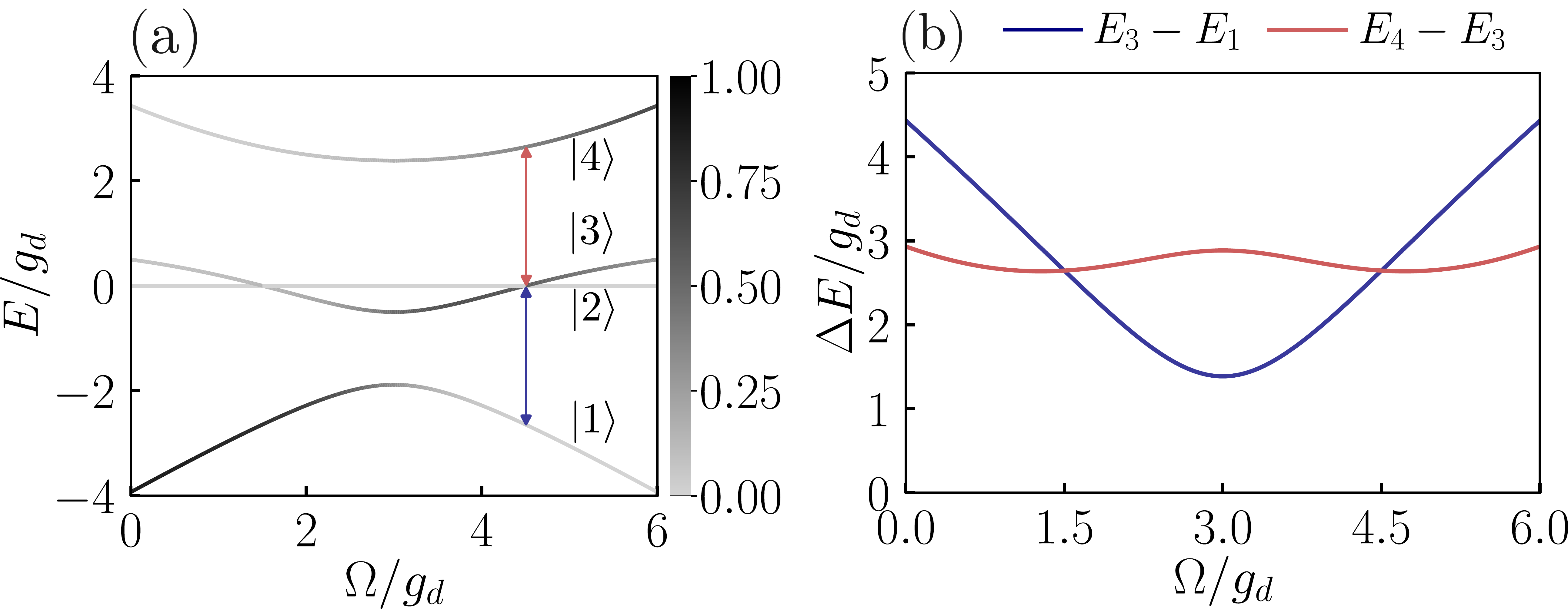}
    \caption{Fig (a) shows the energy spectrum and (b) shows the energy difference between eigenstates $\ket{1}$, $\ket{3}$ and $\ket{4}$ of the Hamiltonian in (\ref{hrot}) for $\beta_z=3$ and $\beta_\perp=2$. The gray color scale in (a) indicates the overlap between the initial state $\ket{\downarrow\downarrow}$ and the energy eigenstates. The eigenstates $\ket{1}$, $\ket{3}$ and $\ket{4}$ can in general be written as a superposition of $\ket{\downarrow\downarrow}$, $\ket{+}$, and $\ket{\uparrow\uparrow}$.}
    \label{fig:espec}
\end{figure}

Interestingly, we observe a sharp dip that appear in $(S_A)_{\rm max}$, which we term as {\em entanglement kinks}, when $\beta_\perp$ is sufficiently large such that the two resonance peaks overlap to form a single peak [see figure~\ref{fig:smax}(d)]. These kinks indicate a sudden change in the two spin dynamics with a slight variation in $\Omega$. For instance, in figure~\ref{fig:smax}(d), the kink appears at $\Omega=4.5g_d$ for $\beta_z=3$ and $\beta_\perp=2$, and the dynamics around this point for a small change in the values of $\Omega$ is shown in figure~\ref{fig:Omegadyn}. At the kink, the population and entanglement dynamics are governed by a single frequency, as shown in figures~\ref{fig:Omegadyn}(b) and \ref{fig:Omegadyn}(e). The emergence of a single frequency in the dynamics at the kink can be understood from the spectrum of $\hat H_\text{rot}$. In particular, the energy spacings between the relevant eigenstates of $\hat H_\text{rot}$ become a multiple of a single frequency, which then governs the dynamics. Based on that, and considering the three relevant states for a pair of $J=1/2$ spins, the criterion for kinks can be obtained as, 
\begin{equation}
\beta_\perp^2=2\left(\beta_z-\frac{\Omega}{g_d}\right)^2-\frac{1}{2},
\label{crit1}
\end{equation}
which indicates an interplay between the magnetic fields and the DDIs.

Now, we particularly focus on the kink appearing in figure~\ref{fig:smax}(d). In figure~\ref{fig:espec}(a), we show the energy eigenvalues of $\hat H_\text{rot}$ as a function of $\Omega$ for $\beta_z=3$ and $\beta_\perp=2$, with the gray scale indicating the overlap of the initial state with each of the energy eigenstates. The horizontal gray line ($E_2$) in figure~\ref{fig:espec}(a) corresponds to the anti-symmetric state $\ket{-}=(\ket{\uparrow\downarrow}-\ket{\downarrow\uparrow})/\sqrt{2}$, which is irrelevant in our case. In figure~\ref{fig:espec}(b), we show the energy differences among the three relevant energy eigenvalues ($E_1$, $E_3$, and $E_4$), with $E_3-E_1$ and $E_4-E_3$ exhibiting two crossings at $\Omega/g_d = 1.5$ and $\Omega/g_d = 4.5$, as expected from equation~\ref{crit1}. At these two crossings, the effective longitudinal magnetic field, $\beta_z-\Omega/g_d$, has the same magnitude but opposite signs for $\beta_z=3$. Therefore, we anticipate that the crossing at $\Omega/g_d = 1.5$ would correspond to a kink if the initial state is $\ket{\uparrow\uparrow}$. It could also be interpreted from the fact that the Hamiltonian is invariant under the transformation $\beta_z-\Omega/g_d\to -(\beta_z-\Omega/g_d)$ and $\hat J_{iz}\to -\hat J_{iz}$, $\hat J_{ix}\to \hat J_{ix}$, and $\hat J_{iy}\to \hat J_{iy}$ with $i\in \{1, 2\}$. Though tedious, the exact analytical results for the dynamics at the kink can be straightforwardly obtained using the kink criterion and the exact eigenstates of $\hat{H}_{\rm rot}$ provided in \ref{kinkeqn}. For the kink whose dynamics is shown in figure~\ref{fig:Omegadyn}(b) and (e), i.e. at $\Omega=4.5g_d$, $\beta_z=3$, and $\beta_\perp=2$, we obtain 
\begin{equation}\label{eq:time_evolved_state_rot_frame_kink}
\begin{split}
    \ket{\psi (t)}_R = &\left[\dfrac{1}{7} \left( 4 + 3\cos \sqrt{7} g_d t \right) - \dfrac{i}{\sqrt{7}} \sin \sqrt{7} g_d t \right] \ket{\downarrow \downarrow} \\
    &+ \left[\dfrac{2\sqrt{2}}{7} \left( \cos \sqrt{7} g_d t - 1 \right) - \dfrac{i\sqrt{2}}{\sqrt{7}} \sin \sqrt{7} g_d t \right] \ket{+} + \dfrac{2}{7}(\cos \sqrt{7} g_d t - 1) \ket{\uparrow \uparrow}.
\end{split}
\end{equation}
Note that the energy spacings of $\hat H_\text{rot}$ becomes a multiple of $\sqrt{7} g_d$ at this kink. The population and entanglement entropy dynamics obtained from equation (\ref{eq:time_evolved_state_rot_frame_kink}) [see \ref{kinkeqn}] are in an excellent agreement with the numerical results shown in figures~\ref{fig:Omegadyn}(b) and \ref{fig:Omegadyn}(e). The population from $\ket{\downarrow \downarrow}$ state periodically transfers to $\ket{+}$ and $\ket{\uparrow \uparrow}$ states [see figure~\ref{fig:Omegadyn}(b)]. The population in $\ket{+}$ predominantly determines the correlation between the spins in this case, but in general, the relative phases between the states $\ket{\downarrow \downarrow}$, $\ket{+}$ and $\ket{\uparrow\uparrow}$ is also a crucial factor \cite{Dhiya_PRA2023}. The population in $|\uparrow\uparrow\rangle$ prevents $S_A$ from attaining the maximum value of one, leading to the kink. Away from the kink, the dynamics involve multiple frequencies [see figures~\ref{fig:Omegadyn}(a) and \ref{fig:Omegadyn}(c) on either side of the kink and \ref{kinkeqn} for analytical estimates], and the quantum interference eventually makes $(S_A)_{\rm max}\sim 1$, in which the populations and relative phases become equally important. Since equation~(\ref{crit1}) is a relation among $\beta_z$, $\beta_\perp$ and $\Omega$, the kinks can also emerge as a function of $\beta_z$ and $\beta_\perp$.

{\em Connection to other spin Hamiltonians.---} The above studies are based on a setup involving two spins interacting via dipole-dipole interactions and placed along the $z$-axis, which give rise to the Hamitlonian in equation~(\ref{hrot}). Noting that the term $\mathbf{\hat{S}}_1 \cdot \mathbf{\hat{S}}_2$ only introduces an overall energy shift in the basis: $\ket{\uparrow \uparrow}, \ \ket{+}$ and $\ket{\downarrow \downarrow}$, the same entanglement kinks and resonances can be equivalently observed in an Ising Hamiltonian with longitudinal and transverse fields, i.e, $\hat{H}_{\rm Ising} = (\beta_z - \Omega/g_d) (\hat{S}_{1z} + \hat{S}_{2z}) + \beta_\perp (\hat{S}_{1x} + \hat{S}_{2x}) - 3 \hat{S}_{1z} \hat{S}_{2z}$. Further, we can introduce the XY model, interms of $\hat H_\text{rot}$ as $\hat{H}_{\rm XY} = \hat H_\text{rot} + 2 \hat{S}_{1z} \hat{S}_{2z}$, which would then have the kink criterion as,

\begin{equation}\label{eq:XY_kinkcrit}
    \beta_\perp^2 = \dfrac{1}{\sqrt{2}} \left[ \dfrac{73}{35} \left( \beta_z - \dfrac{\Omega}{g_d} \right)^2 - \dfrac{1}{12} \right].    
\end{equation}
The Hamiltonians of both $\hat{H}_{\rm Ising}$ and $\hat{H}_{\rm XY}$ are easily accessible in ultra cold Rydberg atom setups \cite{10.1063/5.0211071}.

\begin{figure}
    \centering
    \includegraphics[width=1\linewidth]{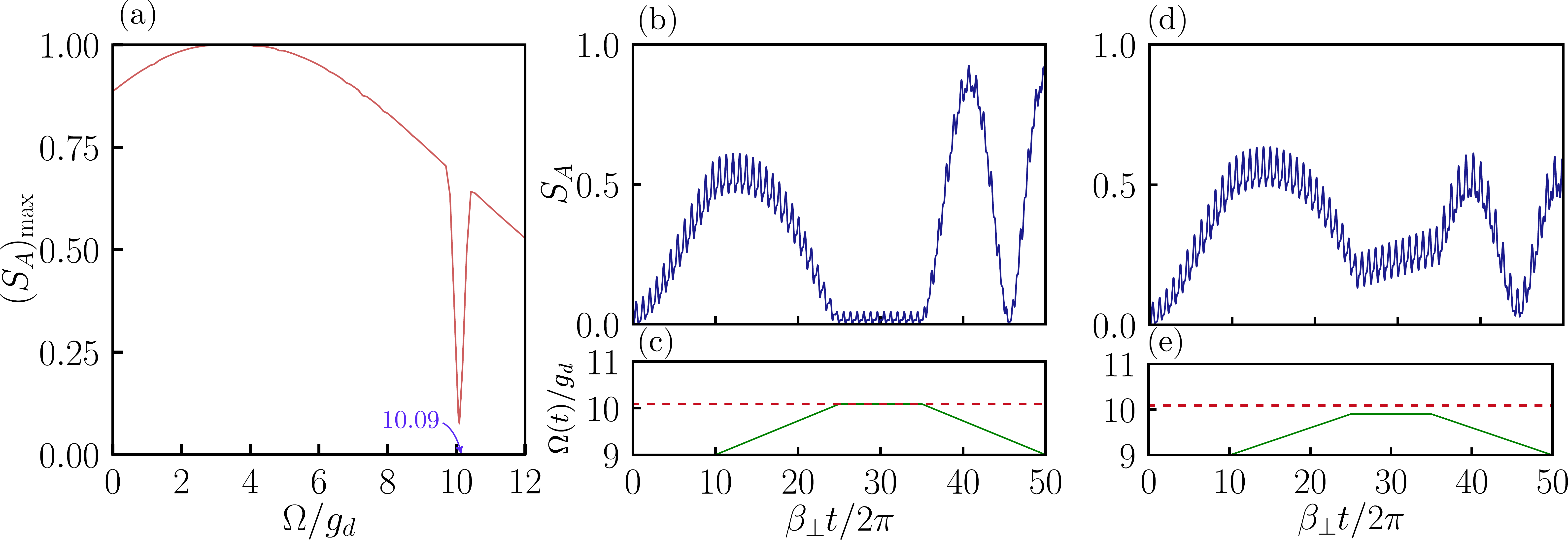}
    \caption{(a) shows the maximum of entanglement entropy as a function of $\Omega/g_d$ for  $\beta_z=3$ and $\beta_\perp=10$ for two spin-$1/2$ particles with dynamics computed over a time period, $T=50(2\pi/g_d\beta_\perp)$. (b) shows the dynamics of entanglement entropy for a linear quench of rotation frequency, $\Omega$, from $\Omega = 9 g_d$ to the kink ($\Omega=10.09 g_d$) and back with a quench rate of $3.63 g_d/T$, as shown in (c). (d) and (e) same as in (b) and (c) respectively, but quenched from $\Omega = 9 g_d$ to $\Omega = 9.9 g_d$ (before the kink) and back with a quench rate of $3g_d/T$. Red dashed line in (c) and (e) shows the kink value, $\Omega=10.09 g_d$.}
    \label{fig:kink_quench_protocol}
\end{figure}

{ \em Significance of the kink to entanglement engineering.---} We now show that by exploiting the kink, we can engineer the entanglement dynamics in a pair of qubits in a controlled way. Using the kink criterion, we identify a kink that has a larger dip in $(S_A)_{\rm max}$ than the one discussed previously, as shown in figure~\ref{fig:kink_quench_protocol}(a) for $\beta_\perp = 10$, $\beta_z = 3$, and $\Omega \simeq 10.09 g_d$. In figure~\ref{fig:kink_quench_protocol}(b), we show the entanglement dynamics for the linear quench of $\Omega(t)$ shown in figure~\ref{fig:kink_quench_protocol}(c). Since the initial value of $\Omega$ is taken slightly away from the kink, a significant entanglement is generated during the initial stage of the dynamics. However, as $\Omega$ is quenched towards the value corresponding to the kink, the entanglement begins to decrease. Interestingly, if the value of $\Omega$ is held constant at the kink, the system maintains the same maximum entanglement it achieved upon reaching that point. In the particular example shown in figure~\ref{fig:kink_quench_protocol}(b), the initial linear quench nearly eliminates entanglement entirely, but in general it depends on the quench rate. The fact that maximum entanglement is preserved at the kink criterion means we can suppress any correlation growth between the qubits, even while they interact with each other. Upon reversing the quench, the correlations are rebuilt in the dynamics. In contrast, as shown in \ref{fig:kink_quench_protocol}(d), a linear quench to a value distinct from that obtained using the kink criterion [see figure~\ref{fig:kink_quench_protocol}(e)] exhibits entanglement growth while keeping $\Omega$ constant at that value, indicating less controlled dynamics.

\subsubsection{$J>1/2$.}
\label{dyjg1-2}

Here, we generalize the discussions on the dynamics of two spins with each having $J>1/2$ and for an initial state, $\ket{-J, -J}$, which is symmetric under exchange of the two spins. Since the dynamics is restricted to the subspace of symmetric states under exchange of the two spins, the relevant values of total angular momentum quantum number are $J_{\text{tot}}=2J, 2J-2,$..., 1 or 0. Despite the energy spectrum getting increasingly complex as $J$ increases, we identify the following resonant transitions from $\ket{-J, -J}$ [see \ref{app:resonance_conditions_gen_J} for details]:
\begin{enumerate}
\item to $|J,J\rangle$ when $\Omega/g_d = \beta_z$, 
\item to $\ket{2J; 2J-1}= (\ket{J, J-1} + \ket{J-1, J})/\sqrt{2}$ when $\Omega/g_d = \beta_z + 3J/(4J-1)$, where $\ket{2J; 2J-1}$ represents a state with total angular momentum quantum number, $J_{\text{tot}}=2J$, and its projection along the $z$-axis is provided by $M_j=2J-1$,
\item to $\ket{2J;-2J+1} = (\ket{-J, -J+1} + \ket{-J+1, -J})/\sqrt{2}$  when $\Omega/g_d = \beta_z + 3J$,
\item to 
\begin{align}
&\cos (\gamma_J/2) \ket{2J; -2J+2} + \sin (\gamma_J/2) \ket{2J-2; -2J+2} \nonumber\\
&= \dfrac{1}{\sqrt{2(4J-1)}} \Big[ \left( \sqrt{2J-1} \cos (\gamma_J/2) + \sqrt{2J} \sin (\gamma_J/2) \right) \left( \ket{-J, -J+2} + \ket{-J+2, -J} \right) \nonumber \\
&\hspace{3cm} + \left( 2\sqrt{J} \cos (\gamma_J/2) - \sqrt{2(2J-1)} \sin (\gamma_J/2) \right) \ket{-J+1, -J+1} \Big]
\label{ts1}
\end{align}
when $\Omega/g_d = \beta_z + (4J-1)/2 + \sqrt{64J^4 - 64J^3 + 36 J^2 - 10 J + 1}/[2(4J-1)]$, with $\gamma_J = \tan^{-1}[3\sqrt{2J(2J-1)}/(8J^2 - 4J - 1)]$. 
\item to 
\begin{align}
&\sin (\gamma_J/2) \ket{2J; -2J+2} - \cos (\gamma_J/2) \ket{2J-2; -2J+2} \nonumber\\
&= \dfrac{1}{\sqrt{2(4J-1)}} \Big[ \left( \sqrt{2J-1} \sin (\gamma_J/2) - \sqrt{2J} \cos (\gamma_J/2) \right) \left( \ket{-J, -J+2} + \ket{-J+2, -J} \right) \nonumber \\
&\hspace{3cm} + \left( 2\sqrt{J} \sin (\gamma_J/2) + \sqrt{2(2J-1)} \cos (\gamma_J/2) \right) \ket{-J+1, -J+1} \Big]
\label{ts2}
\end{align}
when $\Omega/g_d = \beta_z + (4J-1)/2 - \sqrt{64J^4 - 64J^3 + 36 J^2 - 10 J + 1}/[2(4J-1)]$. 
\item to 
\begin{align}
&\cos (\gamma_J/2) \ket{2J; 2J-2} + \sin (\gamma_J/2) \ket{2J-2; 2J-2} \nonumber\\
&= \dfrac{1}{\sqrt{2(4J-1)}} \Big[ \left( \sqrt{2J-1} \cos (\gamma_J/2) + \sqrt{2J} \sin (\gamma_J/2) \right) \left( \ket{J,J-2} + \ket{J-2, J} \right) \nonumber \\
&\hspace{3cm} + \left( 2\sqrt{J} \cos (\gamma_J/2) - \sqrt{2(2J-1)} \sin (\gamma_J/2) \right) \ket{J-1, J-1} \Big]
\label{ts3}
\end{align}
when $\Omega/g_d = \beta_z + (4J-1)/[2(2J-1)] + \sqrt{64J^4 - 64J^3 + 36 J^2 - 10 J + 1}/[2(2J-1)(4J-1)]$. 
\item to 
\begin{align}
&\sin (\gamma_J/2) \ket{2J; 2J-2} - \cos (\gamma_J/2) \ket{2J-2; 2J-2} \nonumber\\
&= \dfrac{1}{\sqrt{2(4J-1)}} \Big[ \left( \sqrt{2J-1} \sin (\gamma_J/2) - \sqrt{2J} \cos (\gamma_J/2) \right) \left( \ket{J, J-2} + \ket{J-2, J} \right) \nonumber \\
&\hspace{3cm} + \left( 2\sqrt{J} \sin (\gamma_J/2) + \sqrt{2(2J-1)} \cos (\gamma_J/2) \right) \ket{J-1, J-1} \Big]
\label{ts4}
\end{align}
when $\Omega/g_d = \beta_z + (4J-1)/[2(2J-1)] - \sqrt{64J^4 - 64J^3 + 36 J^2 - 10 J + 1}/[2(2J-1)(4J-1)]$. 
\end{enumerate}

As $J$ increases, the first resonance (i) becomes extremely narrow as a function of $\Omega$ or $\beta_z$ for small values of $\beta_\perp$. It occurs due to the higher-order nature of the transition between $\ket{-J, -J}$ and $\ket{J, J}$; for instance, it is a fourth-order transition when $J=1$. The second resonance (ii) also involves a higher-order process except when $J=1/2$, which we have discussed above. The maximum entanglement achieved under the resonance (ii) is when the spins attain the state, $\ket{2J;2J-1}$ and is $\log_{2J+1}2$. The resonance (iii) is a direct (first-order) transition and the maximum entanglement attained is again $\log_{2J+1}2$ upon fully populating the state $\ket{2J;-2J+1}$. The resonances (iv) and (v) are second-order in nature, irrespective of the values of $J$, whereas the nature of resonances (vi) and (vii) depends on the value of $J$. The entanglement entropy of the transitioned states [see equations~(\ref{ts1})-(\ref{ts4})] in resonances (iv) to (vii) can be written as, $S_A = -2\lambda \log_{2J+1} \lambda - (1-2\lambda) \log_{2J+1} (1-2\lambda)$, where $\lambda = (1/4) + (4J-1)/[4(64 J^4 - 64 J^3 + 36 J^2 - 10J +1)^{1/2}]$ for (iv) and (vi) and $\lambda = (1/4) - (4J-1)/[4(64 J^4 - 64 J^3 + 36 J^2 - 10J +1)^{1/2}]$ for (v) and (vii). Note that in the spin-1 case, resonances (iv) and (v)  overlap with (vi) and (vii), respectively.

\begin{figure}
    \centering
    \includegraphics[width=1
    \linewidth]{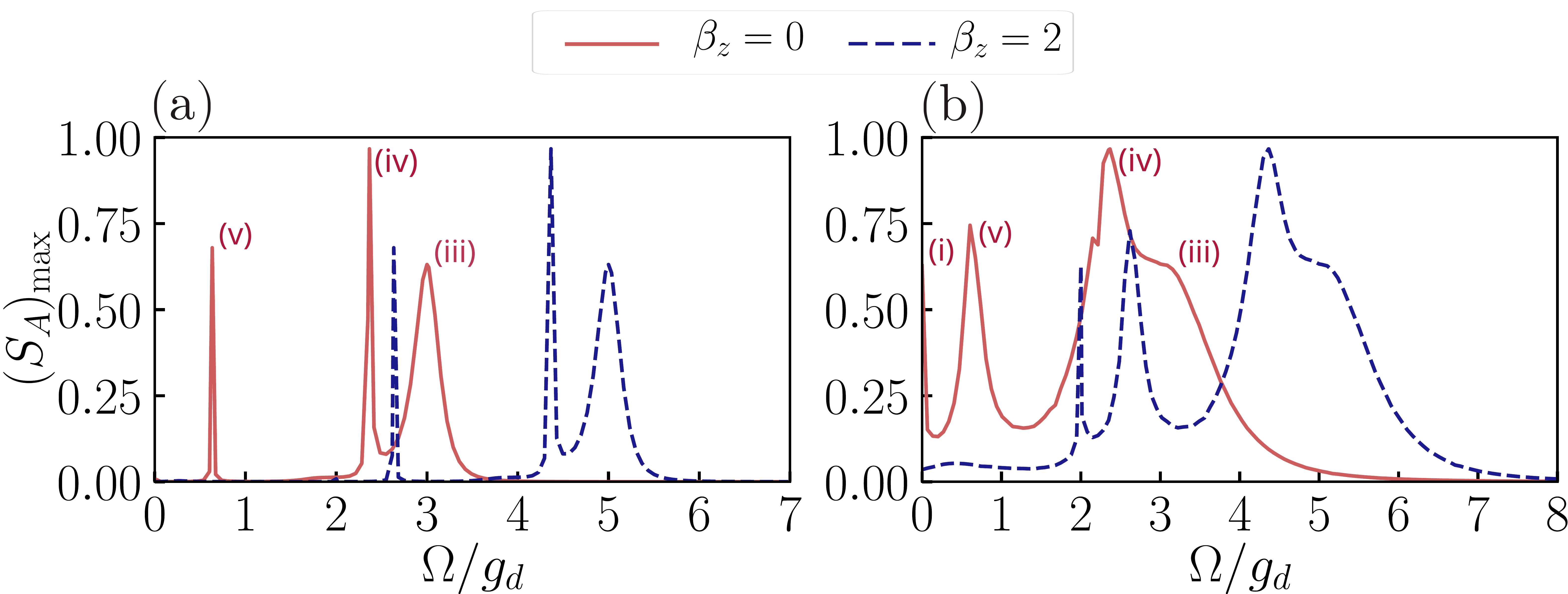}
    \caption{Maximum of entanglement entropy as a function of $\Omega$ for different $\beta_\perp$ and $\beta_z$ for two spin-1 particles. (a) and (b) show the results for $\beta_\perp=$ 0.1 and 0.5, respectively and the dynamics is computed over a time period, $T = 15 (2\pi/g_d \beta_\perp)$.}
    \label{fig:sp1}
\end{figure}
\begin{figure}
    \centering
    \includegraphics[width=1\linewidth]{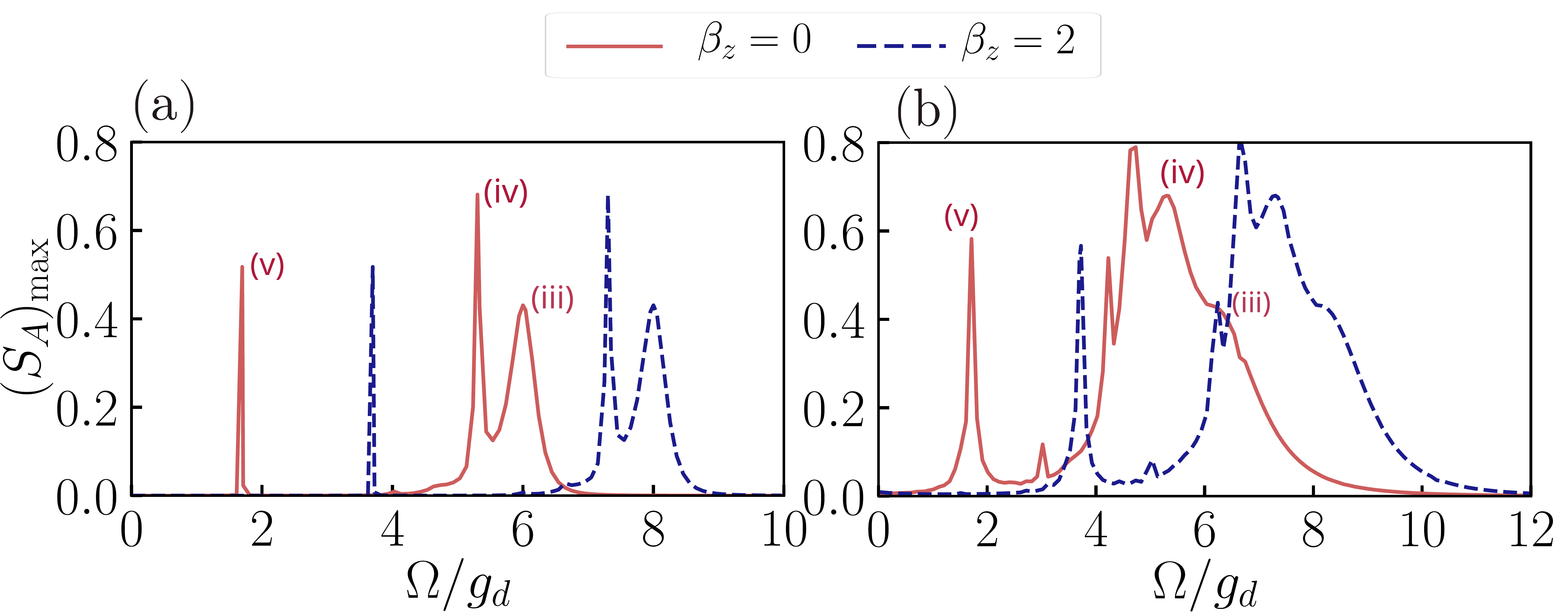}
    \caption{Maximum of entanglement entropy as a function of $\Omega$ for different $\beta_\perp$ and $\beta_z$ for two spin-2 particles. (a) and (b) show the results for $\beta_\perp=$ 0.1 and 0.5 respectively and the dynamics is computed over a time period, $T = 15 (2\pi/g_d \beta_\perp)$.}
    \label{fig:sp2}
\end{figure}

The numerical results for maximum entanglement attained during the dynamics from the initial state $\ket{-1, -1}$ in a pair of spin-1 particles are shown in figure~\ref{fig:sp1}. For smaller $\beta_\perp$ [figure~\ref{fig:sp1}(a)], we can identify three resonances, which are of first and second-order in nature. As $\beta_\perp$ increases, the resonances become broader, allowing the resonance (i), which is fourth-order, to be resolved [see figure~\ref{fig:sp1}(b)] over a period of $T=15 (2\pi/g_d \beta_\perp)$. The numerically calculated peak value of the $(S_A)_{\rm max}$ is in agreement with that of the transitioned state for the resonances (iii) and (iv) in figure~\ref{fig:sp1}. However,  the numerically obtained peak value for resonance (v) is slightly larger than that of the transitioned state in equation~(\ref{ts2}). In this case, the peak value corresponds to a state which is a superposition of $\ket{-1, -1}$ and the state in equation~(\ref{ts2}) with $J=1$, with populations of 0.3 and 0.7, respectively. Similar results are obtained for $J=2$, as shown in figure~\ref{fig:sp2}, where all lower order resonances are captured in the dynamics over a period of $T=15 (2\pi/g_d \beta_\perp)$.

\section{Weak dipolar regime}
\label{wir}
Finally, we briefly discuss the weakly interacting regime, where the dipolar interactions are weak compared to the Zeeman shifts. It is particularly relevant in BEC experiments using chromium \cite{TilmanPfau_Chromium_BEC_2005PRL, Bruno_Laburthe_Chromium_PRA2008}, erbium \cite{Francesca_Erbium_BEC_PRL2012}, and dysprosium \cite{Tuning_DDI_Lev}, where the dipole-dipole interactions can be very weak compared to the Zeeman energies, at least three orders of magnitude smaller. In those cases, the effect of DDIs is to introduce position-dependent energy shifts in the eigenstates of the non-interacting Hamiltonian which, at leading order, is given by the expectation value of the dipolar term, 
\begin{equation}\label{eq:V_ddi_exp_val}
    V_{dd} (\vec{r}) = \langle \hat{V}_{dd} (\vec{r}) \rangle = \dfrac{\mu_0}{4\pi}\dfrac{\langle \hat{\vec{\mu}}_1 \rangle \cdot \langle \hat{\vec{\mu}}_2 \rangle - 3 (\langle \hat{\vec{\mu}}_1 \rangle \cdot \hat{r}) (\langle \hat{\vec{\mu}}_2 \rangle \cdot \hat{r})}{r^3}
\end{equation}
These energy shifts play a crucial role in condensate physics \cite{Lahaye_2009, BARANOV200871}, when they are comparable or even dominant to other energy scales in the system. 

As discussed in Section.~\ref{sdyn}, the dipole moment precesses about the instantaneous direction of the resultant magnetic field with angular frequency $\omega^\prime$ [see \ref{app:dipole_moment_gen_stretched_state} for details of the calculation]. Considering both spins are initialized in the stretched state along an axis forming an angle $\theta_0$ with the $z$-axis, the dynamics of the individual dipole moments is then given by:
\begin{equation}\label{eq:dipole_moment_dyn_gen_lowest_state_main}
    \langle \vec{\mu} \rangle (t) = \mu \cos (\theta_0 - \theta_B) \hat{e} (t) + \mu \sin (\theta_0 - \theta_B) \left[ \cos \omega^\prime t \ \hat{\theta} (t) + \sin \omega^\prime t \ \hat{\varphi} (t) \right],
\end{equation}
 where $\hat{e} (t) = (\sin \theta_B \cos (\Omega t), \sin \theta_B \sin (\Omega t), \cos \theta_B)$ is the unit vector along the effective magnetic field in the lab frame, $\hat{\theta} = (d\hat{e}(t)/d\theta_B)/|d\hat{e}(t)/d\theta_B|$ and $\hat{\varphi} = (d\hat{e}(t)/d(\Omega t))/|d\hat{e}(t)/d(\Omega t)|$ are unit vectors in the direction of the derivatives of $\hat{e}(t)$ with respect to its polar ($\theta_B$) and azimuthal ($\Omega t$) angles, and $\mu = g_J \mu_B (J-n)$ is the magnitude of the dipole moment. For the lowest stretched states, we have $n=0$. In equation (\ref{eq:V_ddi_exp_val}), the first term in $V_{dd} (\vec{r})$ becomes time-independent and equal to $\mu^2$, as the dipole moments remain parallel to each other at all times. The second term contains two oscillating components with frequencies $\omega^\prime$ and $\Omega$. When the dynamics associated with $\omega^\prime$ and $\Omega$ occur at much faster rates compared to the timescale of the DDI strengths and other energy scales, only the time-averaged DDI is significant, which is given by [see \ref{app:DDI_WIR} for details]
\begin{equation}\label{eq:tilman-pfau-generalization-genOmega}
    \overline{V}_{dd}(\vec{r}) = \dfrac{\mu_0 \mu^2}{4\pi r^3} \left(\dfrac{3\cos^2 (\theta_0 - \theta_B) - 1}{2} \right) (1 - 3\cos^2 \theta^\prime) \left( \dfrac{3 \cos^2 \theta_B - 1}{2} \right),
\end{equation}
where we have used the spherical polar coordinates, $\vec{r}=(r, \theta^\prime, \phi^\prime)$. The potential in equation~(\ref{eq:tilman-pfau-generalization-genOmega}) is independent of the azimuthal angle $\phi'$. In the adiabatic limit, i.e. for $\Omega \ll \omega^\prime$, $\theta_B \rightarrow \phi_0$, the direction of the magnetic field at $t=0$. In that case, when $\theta_0 = \phi_0$, we retrieve the case discussed in Refs.~\cite{Tuning_DDI_Tilman_Pfau, Tuning_DDI_Lev}, where the tuning of DDI in a dipolar BEC by means of rotating fields is demonstrated.

\section{Summary and outlook}
\label{suou}

In summary, we studied the dynamics of a single spin and two spins in both static and rotating magnetic fields. In each case, we derived several resonance conditions that characterize coherent and periodic oscillations between different states, including stretched states and entangled states. In the two-atom case, the entanglement generated by the dipolar interactions exhibits a non-trivial dependence on the strength of the magnetic fields. Strikingly, we found entanglement resonances and kinks, along with the criteria for their existence. The kink is characterized by sinusoidal oscillations in entanglement and population dynamics, exhibiting a single frequency with a constant amplitude. These features can be utilized to engineer entanglement dynamics in a highly controlled manner, for instance, through a quench protocol as we have demonstrated.

The entanglement resonances and kinks we found may have promising applications in quantum technologies. Notably, the sharp features of the kinks could be utilized for quantum sensing. These studies can be extended by considering different relative orientations of the two spins with respect to the static field. Another interesting question to investigate is the possibility of observing kinks in spins with $J>1/2$ and in configurations involving more than two spins. In both scenarios, the larger Hilbert space introduces greater complexity in identifying the kinks. Additionally, we could extend this study by incorporating the motional degrees of freedom of the spins and examining the effects of their coupling via DDIs. Other potential avenues for exploration include studying entanglement generation through Landau-Zener sweeps \cite{Dhiya_PRA2023} of magnetic fields, investigating Krylov complexity \cite{Siddharth_PRD2025}, and many-body physics. 


\section*{Acknowledgements}
This work was supported by ``An initiative under the National Quantum Mission (NQM) of Department of Science and Technology (DST)," Government of India. We thank National Supercomputing Mission (NSM) for providing computing resources of ``PARAM Brahma" at IISER Pune, which is implemented by C-DAC and supported by the Ministry of Electronics and Information Technology (MeitY) and Department of Science and Technology (DST), Government of India. We further acknowledge DST-SERB for Swarnajayanti fellowship File No. SB/SJF/2020-21/19, and the MATRICS grant (MTR/2022/000454) from SERB, Government of India and National Mission on Interdisciplinary Cyber-Physical Systems (NM-ICPS) of the Department of Science and Technology, Government of India, through the I-HUB Quantum Technology Foundation, Pune, India. S.S. and N.S. acknowledge funding support from the Senior Research Fellowship (SRF) and Junior Research Fellowship (JRF) respectively, awarded by the University Grants Commission (UGC), India.

\appendix

\section{Dynamics of the dipole moment} 
\label{app:dipole_moment_gen_stretched_state}

In this section, we consider initial states that correspond to a Zeeman sublevel along a quantization axis forming an angle $\theta_0$ with the positive $z$-axis. We assume that this quantization axis is in the same plane as the $z$-axis and the initial direction of the magnetic field, so that the azimuthal angle may be set to 0. It is convenient to represent these states in terms of the $2J$ non-interacting spin $1/2$ particles. We denote the spin-up and down states along this axis for a spin-1/2 particle by $\ket{\uparrow^\prime}$ and $\ket{\downarrow^\prime}$ respectively, which can be obtained by rotating the spin states along the $\hat{z}$-axis by $\theta_0$ about the $\hat{y}$-axis. In terms of the states along the $z$-axis, we then have 
\begin{align}
    \ket{\uparrow^\prime} &= \cos (\theta_0/2) \ket{\uparrow} + \sin (\theta_0/2) \ket{\downarrow} \\
    \ket{\downarrow^\prime} &= -\sin (\theta_0/2) \ket{\uparrow} + \cos (\theta_0/2) \ket{\downarrow}
\end{align}
Then, the $\ket{m_j^\prime = -J + n}$ state of the spin-$J$ particle along the new quantization axis is represented by a symmetric superposition of product states where $n$ of the $2J$ spin 1/2 particles are initialized in $\ket{\uparrow^\prime}$ while the remaining $2J-n$ are initialized in $\ket{\downarrow^\prime}$. There are $^{2J}C_n$ such states in the superposition, resulting in an overall normalization factor of $1/\sqrt{^{2J}C_n}$. We are now interested in calculating the dynamics of the dipole moment component of these initial states, for which we make use of the expectation value of the various spin operators as a function of time. For a spin-1/2 particle in initial state $\ket{\uparrow^\prime}$ or $\ket{\downarrow^\prime}$, the state evolves with time [using equation (\ref{eq:noninteracting-spin-1/2-rotating-frame})] as:

\begin{align}
    \ket{\uparrow^{\prime} (t)} &= e^{-i\Omega t/2} \left(\cos \dfrac{\theta_0}{2} \cos \left( \dfrac{\omega^\prime t}{2}\right) - i \sin \left( \dfrac{\omega^\prime t}{2} \right) \left[ \cos \theta_B \cos \dfrac{\theta_0}{2} + \sin \theta_B \sin \dfrac{\theta_0}{2} \right]\right) \ket{\uparrow} \nonumber \\ 
   &\hspace{0.5cm} + e^{i\Omega t/2} \left( \sin \dfrac{\theta_0}{2} \cos \left( \dfrac{\omega^\prime t}{2}\right) + i \sin \left( \dfrac{\omega^\prime t}{2} \right) \left[ \cos \theta_B \sin \dfrac{\theta_0}{2} - \sin \theta_B \cos \dfrac{\theta_0}{2} \right]\right)\ket{\downarrow} \\
    \ket{\downarrow^\prime (t)} &= 
    e^{-i\Omega t/2} \left(-\sin \dfrac{\theta_0}{2} \cos \left( \dfrac{\omega^\prime t}{2}\right) - i \sin \left( \dfrac{\omega^\prime t}{2} \right) \left[ - \cos \theta_B \sin \dfrac{\theta_0}{2} + \sin \theta_B \cos \dfrac{\theta_0}{2} \right]\right) \ket{\uparrow} \nonumber \\ 
   &\hspace{0.5cm} + e^{i\Omega t/2} \left( \cos \dfrac{\theta_0}{2} \cos \left( \dfrac{\omega^\prime t}{2}\right) + i \sin \left( \dfrac{\omega^\prime t}{2} \right) \left[ \cos \theta_B \cos \dfrac{\theta_0}{2} + \sin \theta_B \sin \dfrac{\theta_0}{2} \right]\right)\ket{\downarrow} \label{eq:time_dynamics_gen_spin_down_state}
\end{align}
where we have substituted for $\cos \theta_B = (\omega_z - \Omega)/\omega^\prime$ and $\sin \theta_B = \omega_\perp/\omega^\prime$. Recall that $\theta_B$ is the direction of the effective magnetic field in the rotating frame, as explained in the main text. The expectation value of the spin operators in these states, as a function of time, can then be obtained as:

\begin{align}
    \langle \hat{\sigma}_x \rangle_{\downarrow^\prime} (t) &= - \cos (\theta_0 - \theta_B) \sin \theta_B \cos \Omega t - \sin (\theta_0 - \theta_B) \left[ \cos \omega^\prime t \cos \theta_B \cos \Omega t - \sin \omega^\prime t \sin \Omega t \right] \label{eq:sigma_x_expect_value_gen} \\
    \langle \hat{\sigma}_y \rangle_{\downarrow^\prime} (t) &= - \cos (\theta_0 - \theta_B) \sin \theta_B \sin \Omega t - \sin (\theta_0 - \theta_B) \left[ \cos \omega^\prime t \cos \theta_B \sin \Omega t + \sin \omega^\prime t \cos \Omega t \right] \label{eq:sigma_y_expect_value_gen} \\
    \langle \hat{\sigma}_z \rangle_{\downarrow^\prime} (t) &= -\cos (\theta_0 - \theta_B) \cos \theta_B + \sin (\theta_0 - \theta_B) \sin \theta_B \cos \omega^\prime t  \label{eq:sigma_z_expect_value_gen}
\end{align}
with $\langle \hat{\sigma}_a \rangle_{\uparrow^\prime} (t) = - \langle \hat{\sigma}_a \rangle_{\downarrow^\prime} (t)$ for $a = x,y,z$. 

We now return to the calculation of the expectation value of the dipole moment operators in the full state, $\ket{\Psi (t)}$. As the state is symmetric under exchange of any two spins and the operators involved can also be decomposed into a symmetric superposition of single-spin operators, it suffices to simply use one of the states from the superposition in $\ket{\Psi (t)}$, say the state where the first $n$ spins are initialized in $\ket{\uparrow^\prime}$ while the remaining $(2J-n)$ are initially in $\ket{\downarrow^\prime}$. Let this state be $\ket{s_1}$, where $\ket{\Psi} = (1/\sqrt{^{2J}C_n})\sum\limits_{i=1}^{^{2J}C_n} \ket{s_i}$. The expectation value of $\hat{J}_a$ ($a = x,y,z$) is then given by:

\begin{equation}
    \begin{split}
        \bra{\Psi (t)} \hat{J}_a \ket{\Psi (t)} &= \dfrac{^{2J}C_n}{2\sqrt{^{2J}C_n}} \sum\limits_{i=1}^{2J} \langle \Psi (t)|\hat{\sigma}^i_a \ket{s_1 (t)} \\
        &= \dfrac{1}{2} \sum\limits_{i=1}^{2J} \sum\limits_{j=1}^{^{2J}C_n} \langle s_j (t)|\hat{\sigma}_a^i|s_1 (t)\rangle \\
        &= \dfrac{1}{2} \sum\limits_{i=1}^{2J} \langle s_1 (t)|\hat{\sigma}_a^i |s_1 (t) \rangle
    \end{split}
\end{equation}
where we use the fact that $\ket{s_j}$ is orthogonal to $\hat{\sigma}_a^i \ket{s_1}$ for $j \neq 1$, as $\ket{s_1}$ and $\ket{s_j}$ differ in the spin state of at least two spins, say at $i_1$ and $i_2$, at least one of which is unaffected by $\hat{\sigma}_a^i$, so that the overlap between the individual spin states vanishes for at least one of the $2J$ spins. Thus, we have

\begin{equation}
    \begin{split}
        \langle \hat{J}_a \rangle (t) &= \dfrac{1}{2} \sum\limits_{i=1}^{2J} \bra{s_1 (t)} \hat{\sigma}_a^i \ket{s_1 (t)} \\
        &= \dfrac{1}{2} \left[ n \langle \hat{\sigma}_a \rangle_{\uparrow^\prime} + (2J-n) \langle \hat{\sigma}_a \rangle_{\downarrow^\prime} \right] \\
        &= (J-n) \langle \hat{\sigma}_a \rangle_{\downarrow^\prime}
    \end{split}
\end{equation}
for $a = x,y,z$. 

The expectation value of the dipole moment operators are then given by $\langle \mu_a \rangle (t) = - g_J \mu_B \langle \hat{J}_a \rangle$. Using equations (\ref{eq:sigma_x_expect_value_gen}) - (\ref{eq:sigma_z_expect_value_gen}), we find that

\begin{equation}\label{eq:dipole_moment_dyn_gen_lowest_state}
    \langle \vec{\mu} \rangle (t) = \mu \cos (\theta_0 - \theta_B) \hat{e} (t) + \mu \sin (\theta_0 - \theta_B) \left[ \cos \omega^\prime t \ \hat{\theta} (t) + \sin \omega^\prime t \ \hat{\varphi} (t) \right],
\end{equation}
where $\hat{e} (t) = (\sin \theta_B \cos (\Omega t), \sin \theta_B \sin (\Omega t), \cos \theta_B)$ is the unit vector along the effective magnetic field in the lab frame, $\hat{\theta} = (d\hat{e}(t)/d\theta_B)/|d\hat{e}(t)/d\theta_B|$ and $\hat{\varphi} = (d\hat{e}(t)/d(\Omega t))/|d\hat{e}(t)/d(\Omega t)|$ are unit vectors in the direction of the derivatives of $\hat{e}(t)$ with respect to its polar ($\theta_B$) and azimuthal ($\Omega t$) angles, and $\mu = g_J \mu_B (J-n)$ is the magnitude of the dipole moment. Note that the negative sign in $\mu$ for $n>J$ indicates that the dipole moment for these states at $t=0$ is opposite to the direction of the effective magnetic field along the quantization axis. 

Equation~(\ref{eq:dipole_moment_dyn_gen_lowest_state}) represents the precession of the dipole moment about the instantaneous effective magnetic field in the lab frame. Note that by setting $n=0$ and $\theta_0 = 0$ or $\phi_0$, we recover the results obtained for the lowest stretched states along the $\hat{z}$-axis and the initial direction of the magnetic field respectively. Similarly, setting $\theta_0 = 0$ and considering all values of $n$ give us the results for the system initially in a given Zeeman sublevel along the $\hat{z}$-axis. The dynamics is clearly identical to the lowest stretched state along the $\hat{z}$-axis, albeit with a reduced dipole moment. 

\section{Other details of dynamics for the lowest stretched states}

In this section, we derive a few more results for the dynamics of the system when it is initialized specifically in the lowest stretched state along a quantization axis making an angle, $\theta_0$, with the $z$-axis. We set the azimuthal angle to be 0 as before. The discussions in the main text pertain to $\theta_0 = 0$ [see section \ref{sec:stretched_zquant}] and $\theta_0 = \phi_0$ [see section \ref{sec:stretched_magnetic_field}]. The state at later times is given by $\ket{\Psi (t)} = \prod\limits_{i=1}^{2J} \ket{\psi_i (t)}$, where $\ket{\psi_i (t)}$ is obtained from equation (\ref{eq:time_dynamics_gen_spin_down_state}). 

\subsection{Survival probability} \label{app:survival_probability_gen}

The survival probability of this initial state is given by

\begin{equation}
    S (t) = |\langle \Psi_0|\Psi(t) \rangle|^2 = \prod\limits_{i=1}^{2J} |\langle \psi_{i,0}|\psi_{i} (t) \rangle|^2 = \left( \left|\langle \psi_{1,0}|\psi_1 (t) \rangle \right|^2 \right)^{2J}
\end{equation}
as all the individual spins are in identical states at all times. We get 

\begin{equation}
    \begin{split}
        \langle \psi_{i,0}|\psi_{i} (t) \rangle &= \cos \left( \dfrac{\omega^\prime t}{2} \right) \cos \left( \dfrac{\Omega t}{2} \right) - \cos \theta_B \sin \left( \dfrac{\omega^\prime t}{2} \right) \sin \left( \dfrac{\Omega t}{2} \right) \\
        & \hspace{0.5cm} + i \Bigg[ \cos \theta_0 \cos \left( \dfrac{\omega^\prime t}{2} \right) \sin \left( \dfrac{\Omega t}{2} \right) + \cos (\theta_0 - \theta_B) \sin \left( \dfrac{\omega^\prime t}{2} \right) \cos \left( \dfrac{\Omega t}{2} \right) \Bigg]
    \end{split}
\end{equation}
and, ultimately, the survival probability as

\begin{align}
    S &= \Bigg[ 1 + \dfrac{1}{2} \left[ \cos \theta_0 \cos (\theta_0 - \theta_B) - \cos \theta_B \right] \sin \omega^\prime t \sin \Omega t - \sin^2 \theta_0 \cos^2 \left( \dfrac{\omega^\prime t}{2} \right) \sin^2 \left( \dfrac{\Omega t}{2} \right) \nonumber \\
    &\hspace{2cm} - \sin^2 \left(\dfrac{\omega^\prime t}{2}\right) \left[ \sin^2 \theta_B \sin^2 \left( \dfrac{\Omega t}{2} \right) + \sin^2 (\theta_0 - \theta_B) \cos^2 \left( \dfrac{\Omega t}{2} \right) \right] \Bigg]^{2J}
\end{align}
Setting $\theta_0 = 0, \phi_0$, we recover the results in the main text. The former gives us 

\begin{equation}
    S = \left[ 1 - \dfrac{\omega_\perp^2}{\omega^{\prime 2}} \sin^2 \left( \dfrac{\omega^\prime t}{2} \right) \right]^{2J}
\end{equation}
while the latter, after some additional simplification, reduces to equation (\ref{St3}) in the main text. 

We can similarly also calculate the survival probability in the rotating frame. The individual spin states in the rotating frame are again obtained from equation (\ref{eq:time_dynamics_gen_spin_down_state}), but this time we discard the phase factors $e^{\pm i\Omega t/2}$ in front of the spin-$\uparrow/\downarrow$ components, which were a result of the transformation back to the lab frame. Repeating the procedure, the survival probability in the rotating frame is obtained as 

\begin{equation}
    S_{\text{rot}} (t) = \left[1 - \sin^2 (\theta_0- \theta_B) \sin^2 \left( \dfrac{\omega^\prime t}{2} \right)\right]^{2J}
\end{equation}
with a minimum survival probability of $S_{\text{rot,min}} = \left[\cos^2 (\theta_0 - \theta_B)\right]^{2J}$. For $\theta_0 = \phi_0$, $S_{\text{rot}} (t)$ is simply the projection of the time-evolved state onto the instantaneous ground state of the system. Note that in the adiabatic limit (i.e. for $\Omega/\omega_z \rightarrow 0$), $S_{\text{rot,min}} \rightarrow 1$, i.e. the system is always in the instantaneous ground state.

\subsection{Population in various sublevels}\label{app:population_levels_gen}

We now obtain the population of the time-evolved state in $\ket{m_j = -J+n}$. As mentioned earlier, this state may be represented as a superposition of all states with $n$ of the $2J$ spins in $\ket{\uparrow}$ and the remaining ($2J-n$) spins in $\ket{\downarrow}$; importantly, we recall that the state is symmetric under exchange of any pair of spins. As $\ket{\Psi (t)}$ is a product state of identical spin states and is also symmetric under exchange of any two spins, it is sufficient to calculate the projection of $\ket{\Psi (t)}$ onto any of the terms in the superposition and simply multiply the result with the total number of the terms in the superposition. Thus, the projection of $\ket{\Psi (t)}$ onto $\ket{-J+n}$ is given by

\begin{equation}
    \langle -J+n|\Psi (t) \rangle = \dfrac{{}^{2J}C_n}{\sqrt{{}^{2J}C_n}} \cdot \left( \langle \uparrow|\psi_1 (t) \rangle \right)^n \left( \langle \downarrow|\psi_1 (t) \rangle \right)^{2J-n}
\end{equation}
where the denominator is the normalization constant in the superposition that constitutes $\ket{-J+n}$, while the numerator is the number of terms in the superposition. Using equation (\ref{eq:time_dynamics_gen_spin_down_state}), the population in $\ket{-J+n}$ is, thus, obtained as:

\begin{equation}
    \begin{split}
        P_{n} (t) &= \left| \langle -J+n|\Psi (t) \rangle \right|^2 = {}^{2J}C_n \left[ p(t) \right]^n [q(t)]^{2J-n}
    \end{split}
\end{equation}
where 

\begin{align}
    p(t) &= \sin^2 \dfrac{\theta_0}{2} \cos^2 \left( \dfrac{\omega^\prime t}{2} \right) + \sin^2 \left( \dfrac{\omega^\prime t}{2} \right) \sin^2 \left(\theta_B - \dfrac{\theta_0}{2} \right) \\
    q(t) &= \cos^2 \dfrac{\theta_0}{2} \cos^2 \left( \dfrac{\omega^\prime t}{2} \right) + \sin^2 \left( \dfrac{\omega^\prime t}{2} \right) \cos^2 \left(\theta_B - \dfrac{\theta_0}{2} \right) 
\end{align}
Setting $\theta_0 = 0$ gives us equations (\ref{eq:p}) and (\ref{eq:q}), respectively, in the main text. We also obtain the maximum population in any of the $n$ or $m_j$ sublevels. Setting $p_{\rm max} = {\rm max} \{\sin^2 (\theta_0/2), \sin^2 (\theta_B - (\theta_0/2))\}$ (i.e. the maximum value of $p(t)$), we find that 

\begin{equation}
    P_{n,\text{max}} = \begin{cases}
        {}^{2J}C_n \left( p_{\rm max} \right)^{n} \left( 1 - p_{\rm max} \right)^{2J-n}, & p_{\rm max} \leq \dfrac{n}{2J} \\
        {}^{2J}C_n \left( \dfrac{n}{2J} \right)^n \left( \dfrac{2J-n}{2J} \right)^{2J-n}, & p_{\rm max} > \dfrac{n}{2J},
    \end{cases}
\end{equation}
which can attain a value of 1, only if $n=0$ or $n=2J$, i.e. it is only possible to completely populate the stretched states. For $\theta_0 = 0$, i.e. when the initial state is taken as $\ket{-J}$, we obtain:

\begin{equation}
    P_{n,\text{max}} = \begin{cases}
        {}^{2J}C_n \left( \dfrac{\omega_\perp^2}{\omega^{\prime 2}} \right)^{n} \left( \dfrac{(\omega_z - \Omega)^2}{\omega^{\prime 2}} \right)^{2J-n}, & \dfrac{\omega_\perp^2}{\omega^{\prime 2}} \leq \dfrac{n}{2J} \\
        {}^{2J}C_n \left( \dfrac{n}{2J} \right)^n \left( \dfrac{2J-n}{2J} \right)^{2J-n}, & \dfrac{\omega_\perp^2}{\omega^{\prime 2}} > \dfrac{n}{2J},
    \end{cases}
\end{equation}
For $\theta_0 = \phi_0$, we get:

\begin{align}
    p(t) &= \sin^2 \dfrac{\phi_0}{2} + \dfrac{\Omega \omega \sin^2 \phi_0}{\omega^{\prime 2}} \sin^2 \left( \dfrac{\omega^\prime t}{2} \right) \\
    q(t) &= \cos^2 \dfrac{\phi_0}{2} - \dfrac{\Omega \omega \sin^2 \phi_0}{\omega^{\prime 2}} \sin^2 \left( \dfrac{\omega^\prime t}{2} \right)
\end{align}
In this case, it can be readily verified from above that for $\Omega = \omega_z (B/B_z) = \omega$, $\omega^\prime = \sqrt{2} \omega [1 - (\omega_z/\omega)]^{1/2} = 2\omega \sin (\phi_0/2)$ and the maximum value of $P_{2J} (t) = 1$, indicating that $\ket{m_j = J}$ will be completely populated periodically. 

\subsection{Spread in the Zeeman sublevels} \label{app:spread_of_state_gen}

We quantify the spread of the state in the Zeeman sublevels (along the $z$-axis) as $\sqrt{\langle \hat{J}_z^2 \rangle - \langle \hat{J}_z \rangle^2}$. Now, $\langle \hat{J}_z \rangle$ is readily obtained as $J \langle \hat{\sigma}_z \rangle$ from equation (\ref{eq:sigma_z_expect_value_gen}) as all the spins are in identical states. For $\langle \hat{J}_z^2 \rangle$, we express it as

\begin{align}
    \langle \hat{J}_z^2 \rangle &= \langle \left( \sum\limits_{i=1}^{2J} \dfrac{\hat{\sigma}_z^i}{2} \right)^2 \rangle = \dfrac{1}{4} \left[ \sum\limits_{i=1}^{2J} (1) + \sum\limits_{i \neq j} \langle \hat{\sigma}_z^i \rangle \langle \hat{\sigma}_z^j \rangle \right] = \dfrac{J}{2} + \dfrac{J(2J-1)}{2} \left(\langle \hat{\sigma}_z \rangle_{\downarrow^\prime} \right)^2
\end{align}
where we used the fact that the total state at all times is a tensor product of identical spin states of the individual spins and that the two spins are non-interacting in order to decompose the two-point correlations into a product of the expectation values of the two spins involved. Thus, we get

\begin{equation}
    \begin{split}
        \Delta m_j = \sqrt{\langle \hat{J}_z^2 \rangle - \langle \hat{J}_z \rangle^2} &= \sqrt{\dfrac{J}{2} \left( 1 - \langle \hat{\sigma}_z \rangle_{\downarrow^\prime}^2 \right)}
    \end{split}
\end{equation}
For the initial state considered in Section \ref{sec:stretched_magnetic_field}, we set $\theta_0 = \phi_0$ above and use equation (\ref{eq:sigma_z_expect_value_gen}) to obtain 

\begin{equation}
    \Delta m_j = \sqrt{\dfrac{J}{2}} \sin \phi_0 \left[ 1 - \dfrac{\omega_\perp^2 \Omega^2}{\omega^{\prime 4}} (1 - \cos \omega^\prime t)^2 + \dfrac{2 \omega_z \Omega}{\omega^{\prime 2}} (1 - \cos \omega^\prime t) \right]^{1/2} 
\end{equation}

\section{Analytical results for the dynamics of two interacting spins at the kink}
\label{kinkeqn}

We obtain analytical results for the population and entanglement entropy dynamics of two interacting spin-1/2 particles in the presence of a rotating magnetic field, specifically at the positions of the entanglement `kinks', described in Section \ref{dyj1-2}. The energy spectrum is obtained by diagonalizing the Hamiltonian in the rotating frame, where we label the energies as $E_i$ ($i = 1,2,3,4$) and the corresponding eigenstates as $\ket{i}$. Initializing the spins in $\ket{\downarrow \downarrow}$, the dynamics takes place in the three-dimensional subspace spanned by $\ket{\downarrow \downarrow}, \ \ket{+}$ and $\ket{\uparrow \uparrow}$, or $\ket{1}, \ \ket{3}$ and $\ket{4}$ in terms of the energy eigenbasis [see Fig.~\ref{fig:espec} in main text]. The state, $\ket{2} = \ket{-} = (\ket{\uparrow \downarrow} - \ket{\downarrow \uparrow})/\sqrt{2}$, completes the Hilbert space of the system and does not play a role in the dynamics. 

The kink in the maximum entanglement entropy appears when $E_3 - E_1 = E_4 - E_3$, resulting in sinusoidal oscillations of the populations and entanglement entropy at a single frequency given by this energy gap. The energy gaps become equal when

\begin{equation}
    \beta_\perp = \sqrt{2 \left(\beta_z - \dfrac{\Omega}{g_d}\right)^2 - \dfrac{1}{2}}.
\end{equation}
The three relevant energy eigenvalues in this case are then given by $E_1 = -\alpha g_d, \ E_3 = 0, \text{ and } E_4 = \alpha g_d$ with $\alpha = \sqrt{(1/4) + 3 [\beta_z - (\Omega/g_d)]^2}$. We set $\Delta = \beta_z - (\Omega/g_d)$ henceforth for brevity. Correspondingly, the three eigenstates are obtained as follows: 

\begin{align}
    \ket{1} &\propto \dfrac{\sqrt{2} \beta_\perp}{2(\Delta - \alpha) + 1} \ket{\downarrow \downarrow} + \ket{+} - \dfrac{\sqrt{2} \beta_\perp}{2(\Delta + \alpha) - 1} \ket{\uparrow \uparrow} \\
    \ket{3} &= \dfrac{\left[ \sqrt{2} \beta_\perp (2\Delta - 1) \ket{\downarrow \downarrow} + (4\Delta^2 - 1) \ket{+} - \sqrt{2} \beta_\perp (2\Delta+1) \ket{\uparrow \uparrow} \right]}{\sqrt{(4\Delta^2 - 1)(12\Delta^2 + 1)}} \\
    \ket{4} &\propto \dfrac{\sqrt{2} \beta_\perp}{2(\Delta + \alpha) + 1} \ket{\downarrow \downarrow} + \ket{+} - \dfrac{\sqrt{2} \beta_\perp}{2(\Delta - \alpha) - 1} \ket{\uparrow \uparrow} 
\end{align}
where $\ket{1}$ and $\ket{4}$ above have been obtained up to an overall normalization constant. 

The kink coincides with the second resonance condition at $(\Omega/g_d) = \beta_z + 3/2$ when $\beta_\perp = 2$. As a result, while away from this point the maximum entanglement entropy in the dynamics is 1, we observe a significant dip at and very close to this point. At exactly $\beta_\perp = 2$, we may use the eigenstates above, and we obtain the time-evolved state in the rotating frame as

\begin{equation}\label{eq:time_evolved_state_rot_frame}
\begin{split}
    \ket{\psi (t)}_R = &\left[\dfrac{1}{7} \left( 4 + 3\cos \alpha g_d t \right) - \dfrac{i}{\sqrt{7}} \sin \alpha g_d t \right] \ket{\downarrow \downarrow} \\
    &+ \left[\dfrac{2\sqrt{2}}{7} \left( \cos \alpha g_d t - 1 \right) - \dfrac{i\sqrt{2}}{\sqrt{7}} \sin \alpha g_d t \right] \ket{+} + \dfrac{2}{7}(\cos \alpha g_d t - 1) \ket{\uparrow \uparrow}
\end{split}
\end{equation}
with $\alpha = \sqrt{7}$ for this particular choice of parameters. The populations in the three states of interest, which remain unaffected by transforming back to the lab frame, are then given by:

\begin{align}
    P_{\downarrow \downarrow} &= \dfrac{1}{49} \left(9 \cos^2 \alpha g_d t + 24 \cos \alpha g_d t + 16 \right) + \dfrac{\sin^2 \alpha g_d t}{7} \\
    P_+ &= \dfrac{8}{49} \left( \cos^2 \alpha g_d t - 2\cos \alpha g_d t + 1 \right) + \dfrac{2}{7} \sin^2 \alpha g_d t \\
    P_{\uparrow \uparrow} &= \dfrac{4}{49} \left( \cos^2 \alpha g_d t - 2\cos \alpha g_d t + 1 \right)
\end{align}
For a general state of the form, $\ket{\psi} = c_{\downarrow \downarrow} \ket{\downarrow \downarrow} + c_+ \ket{+} + c_{\uparrow \uparrow} \ket{\uparrow \uparrow}$, the entanglement entropy can be calculated as $S_A = - \lambda_+ \log_2 \lambda_+ - \lambda_- \log_2 \lambda_-$, where $\lambda_\pm$ are the eigenvalues of the reduced density matrix of one of the qubits and is given by,

\begin{equation}
    \begin{split}
        \lambda_\pm &= \dfrac{1}{2} \pm \dfrac{\sqrt{(P_{\downarrow \downarrow} - P_{\uparrow \uparrow})^2 + 2 P_+ (P_{\downarrow \downarrow} + P_{\uparrow \uparrow}) + 2 (c_+^2 c_{\downarrow \downarrow}^* c_{\uparrow \uparrow}^* + c_+^{*2} c_{\downarrow \downarrow} c_{\uparrow \uparrow})}}{2} 
    \end{split}
\end{equation}
where $P_k = |c_k|^2$ with $k = \downarrow \downarrow, +, \uparrow \uparrow$. It can be readily verified from above that the eigenvalues of the reduced density matrix, and consequently the entanglement entropy, are identical in the lab frame and the rotating frame. Thus, using equation (\ref{eq:time_evolved_state_rot_frame}), we obtain:

\begin{equation}
    \lambda_{\pm} = \dfrac{1}{2} \pm \dfrac{\sqrt{2189 + 624 \cos \alpha g_d t - 600 \cos^2 \alpha g_d t + 176 \cos^3 \alpha g_d t + 12 \cos^4 \alpha g_d t}}{98} .
\end{equation}
The population dynamics and the resulting entanglement entropy are in perfect agreement with the numerical results for this case presented in figure~\ref{fig:Omegadyn} (b) and (e) in the main text. 

Away from the kinks, the oscillation frequencies of the populations are different, as shown in figures~\ref{fig:Omegadyn}(a, c, d, f). Analytically, for $\Omega/g_d = 4.5 - \delta$, we may obtain corrections to the time-evolved state, $\ket{\psi (t)}_R$, in the rotating frame for small $\delta$ using perturbation theory. Incorporating only the lowest-order corrections to the energies (given by $\Delta E_i = \bra{i}\delta (\hat{S}_{1z} + \hat{S}_{2z})\ket{i}$, for $i = 1,3,4$), the state at later times is obtained as (upto global phase factors):

\begin{equation}
    \begin{split}
        \ket{\psi (t)}_R &= \Bigg[ \dfrac{3 \cos \left[\left(1 - \dfrac{3\delta}{4} \right) \sqrt{7} g_d t \right] + 4 \cos \dfrac{9\delta g_d t}{14}}{7} \\
        &\hspace{1cm} - \dfrac{i}{7} \left[ \sqrt{7} \sin \left[\left(1 - \dfrac{3\delta}{4} \right) \sqrt{7} g_d t \right] - 4 \sin \dfrac{9\delta g_d t}{14} \right] \Bigg] \ket{\downarrow \downarrow} \\
        &\hspace{1cm} + \dfrac{\sqrt{2}}{7} \Bigg[ 2\cos \left[\left(1 - \dfrac{3\delta}{4} \right) \sqrt{7} g_d t \right] - 2\cos \dfrac{9\delta g_d t}{14} \\
        &\hspace{1cm} - i \left( \sqrt{7} \sin \left[\left(1 - \dfrac{3\delta}{4} \right) \sqrt{7} g_d t \right] + 2 \sin \dfrac{9\delta g_d t}{14} \right) \Bigg] \ket{+} \\
        &+ \dfrac{2}{7} \left[ \cos \left[\left(1 - \dfrac{3\delta}{4} \right) \sqrt{7} g_d t \right] - \cos \dfrac{9\delta g_d t}{14} - i \sin \dfrac{9 \delta g_d t}{14} \right] \ket{\uparrow \uparrow}
    \end{split}
\end{equation}
The populations obtained from the above are estimates that we expect to be accurate up to corrections of the order $\mathcal{O}(\delta)$. Further corrections may be obtained by including first order corrections to the states, though this is not necessary to understand the temporal behaviour of the entanglement entropy. Even for a small deviation from this kink, say $\delta = 0.001$ $(\text{or } \Omega/g_d = 4.499)$, the entanglement entropy reaches a maximum of $S_A \sim 0.99$, albeit after long times. This can be understood in terms of the dynamics being dictated by multiple frequencies whose difference is very small and thus, the overall period of the dynamics is much longer. As a result, due to the quantum interference arising from the offset of the energy differences which influences both the populations and the relative phases, the maximum of $S_A\sim 1$. In contrast, at the kink, the dynamics strictly involves only a single frequency and is perfectly periodic, with $(S_A)_{\rm max} \sim 0.75$ only.

\section{Resonance conditions for two interacting spin-$J$ particles} \label{app:resonance_conditions_gen_J}

We analytically obtain the resonance conditions (i) - (vii) listed in section \ref{dyjg1-2} where we observe spikes in the maximum entanglement entropy at small $\beta_\perp$. We first obtain the energies of the relevant states of the Hamiltonian in the rotating frame, $\hat{H}_{\text{rot}}$, in the absence of $\beta_\perp$. As the Hamiltonian commutes with the total magnetization, $\hat{J}_z = \hat{J}_{1z} + \hat{J}_{2z}$, the eigenstates of the Hamiltonian have a fixed $M_j = m_{j_1} + m_{j_2}$ value. The Hamiltonian is further symmetric with respect to exchanging the two spins. The eigenstates of the Hamiltonian will also, thus, be either symmetric or antisymmetric with respect to exchange of the two spins. As our initial state, $\ket{-J, -J}$, is symmetric with respect to exchange of the spins, the dynamics takes place only in the symmetric subspace. Writing $\hat{\mathbf{J}} = \hat{\mathbf{J}}_1 + \hat{\mathbf{J}}_2$, the dynamics then only involves states with $J_{\text{tot}} = 2J, 2J-2, \hdots $. 

Due to the mixing of states with different $J_{\text{tot}}$, it is difficult to analytically obtain all the relevant eigenstates of $\hat{H}_{\text{rot}}$ even with $\beta_\perp = 0$. However, as the eigenstates have fixed $M_j$, a few of them can be obtained for any general $J$, which is what we focus on. The states with $M_j = \pm 2J, \ \pm (2J - 1)$ only involve the states with $J_{\text{tot}} = 2J$, i.e. $\ket{2J; \pm 2J} = \ket{\pm J, \pm J}$ and $\ket{2J; \pm (2J-1)} = (\ket{\pm J, \pm (J-1)} + \ket{\pm (J-1), \pm J})/\sqrt{2}$ and their respective energies are readily obtained as $E_{\pm 2J} = [\pm 2J\Delta - 2J^2]g_d$ and $E_{\pm (2J-1)} = [\pm (2J-1)\Delta - 2J^2 + 3J] g_d$, where we set $\Delta = (\beta_z - \Omega/g_d)$. The eigenstates with $M_j = \pm (2J-2)$ also only involve the states $\ket{2J; \pm (2J-2)}$ and $\ket{2J-2; \pm (2J-2)}$ and the exact eigenstates and energies may be obtained by diagonalizing $\hat{H}_{\text{rot}}$ (at $\beta_\perp = 0$) in this two-dimensional subspace. The eigenstates are then obtained as:

\begin{align}
        &\ket{M_j = \pm (2J-2)}_+ = \cos (\gamma_J/2) \ket{2J; \pm (2J-2)} + \sin (\gamma_J/2) \ket{2J-2; \pm (2J-2)} \nonumber \\
        &= \Big[ \left( \dfrac{\sqrt{2J-1} \cos (\gamma_J/2) + \sqrt{2J} \sin (\gamma_J/2)}{\sqrt{2(4J-1)}} \right) \left( \ket{\pm J, \pm (J-2)} + \ket{\pm (J-2), \pm J} \right) \nonumber \\
        &\hspace{2cm} + \left( \dfrac{2\sqrt{J} \cos (\gamma_J/2) - \sqrt{2(2J-1)} \sin (\gamma_J/2)}{\sqrt{2(4J-1)}} \right) \ket{\pm (J-1), \pm (J-1)} \Big] \\
        &\ket{M_j = \pm (2J-2)}_- = \sin (\gamma_J/2) \ket{2J; \pm (2J-2)} - \cos (\gamma_J/2) \ket{2J-2; \pm (2J-2)} \nonumber \\
        &= \Big[ \left( \dfrac{\sqrt{2J-1} \sin (\gamma_J/2) - \sqrt{2J} \cos (\gamma_J/2)}{\sqrt{2(4J-1)}} \right) \left( \ket{\pm J, \pm (J-2)} + \ket{\pm (J-2), \pm J} \right) \nonumber \\
        &\hspace{2cm} + \left( \dfrac{2\sqrt{J} \sin (\gamma_J/2) + \sqrt{2(2J-1)} \cos (\gamma_J/2)}{\sqrt{2(4J-1)}} \right) \ket{\pm (J-1), \pm (J-1)} \Big]
    \end{align}
where $\gamma_J = \tan^{-1} [3\sqrt{2J(2J-1)}/(8J^2 - 4J - 1)]$. The energies of $\ket{M_j = \pm (2J-2)}_{\pm}$ are given by

\begin{align}
    E_{2J-2,\pm} &= \left[(2J-2) \Delta - \dfrac{8J^3 - 18J^2 + 8J - 1}{4J-1} \pm \dfrac{\sqrt{64J^4 - 64J^3 + 36J^2 - 10 J + 1}}{4J-1}\right] g_d \\
    E_{-2J+2,\pm} &= \left[- (2J-2) \Delta - \dfrac{8J^3 - 18J^2 + 8J - 1}{4J-1} \pm \dfrac{\sqrt{64J^4 - 64J^3 + 36J^2 - 10 J + 1}}{4J-1}\right] g_d
\end{align}
The resonance conditions for $\Omega/g_d$ in terms of $\beta_z$ are then obtained by equating the energies of $\ket{-J, -J}$ and the target states above. At finite but small $\beta_\perp$, these energy crossings turn into avoided energy crossings with the energy gap proportional to $\beta_\perp^n$, where $n$ is the order of coupling between the two states involved in the crossing. The dynamics in the vicinity of the avoided crossing resembles that of an effective two-level system effectively involving only these two states, resulting in the observed spikes in entanglement entropy as the target state is completely populated (if the target state is already entangled), or more generally as the two states are superposed in the dynamics. For instance, in the case of transitions between $\ket{-J, -J}$ and $\ket{M_j = -2J+2}_\pm$ (corresponding to resonance conditions (iv) and (v) in the main text), the entanglement entropy of the final state is given by $S_A = -2\lambda \log_{2J+1} \lambda - (1-2\lambda) \log_{2J+1} (1-2\lambda)$, where $\lambda = \lambda_\pm$ for $\ket{-2J+2}_{\pm}$ respectively, with $\lambda_{\pm} = (1/4) \pm (4J-1)/[4(64J^4 - 64J^3 + 36 J^2 - 10J + 1)^{1/2}]$. During the dynamics, assuming a population $p$ in $\ket{-J, -J}$ and $(1-p)$ in $\ket{-2J+2}_{\pm}$, we find the entanglement entropy given by $S_A^{\pm} = - \lambda_1^\pm \log_{2J+1} \lambda_1^\pm - \lambda_2^\pm \log_{2J+1} \lambda_2^\pm - (1-\lambda_1^\pm - \lambda_2^\pm) \log_{2J+1} (1 - \lambda_1^\pm - \lambda_2^\pm)$ where

\begin{align}
    \lambda_1^\pm &= \dfrac{p+2(1-p)\lambda_\pm}{2} + \dfrac{\sqrt{p^2 + 4p(1-p)\lambda_\pm}}{2} \\
    \lambda_2^\pm &= \dfrac{p+2(1-p)\lambda_\pm}{2} - \dfrac{\sqrt{p^2 + 4p(1-p)\lambda_\pm}}{2}
\end{align}
For $J=1$, we find that in the case of $\ket{-2J+2}_+$, the entanglement entropy is maximized for $p=0$ i.e. when the population is completely transferred to the target state. For $\ket{-2J+2}_-$, the entanglement entropy is maximized when $p \sim 0.3$. 

Note that for $J=1/2$, all the symmetric states have $J_{\text{tot}} = 1$, while for $J = 1$, the symmetric states have $J_{\text{tot}} = 2, 0$, so that in these cases, we have obtained all the resonance conditions with our analysis above. For larger $J$, we note that there are other resonance conditions besides the one listed above, though at small $\beta_\perp$, the transitions are at least third-order in $\beta_\perp$ and the corresponding dynamics, thus, takes place at longer timescales (by at least one order of magnitude). 

\section{Time-averaged dipole-dipole interactions in the weakly interacting regime}\label{app:DDI_WIR}

We consider very weak dipole-dipole interactions (DDI) such that they may be treated as a perturbation to the noninteracting Hamiltonian and introduce position-dependent shifts to the energies. Initializing the two spins in identical initial states given by the Zeeman sublevels along a quantization axis that makes an angle, $\theta_0$, with the $z$-axis (we assume the azimuthal angle is 0), the expectation value of the DDI is then given by

\begin{equation}
    V_{dd} (\vec{r},t) = \langle \hat{V}_{dd} (\vec{r}) \rangle (t) = \dfrac{\mu_0}{4\pi}\dfrac{\langle \hat{\vec{\mu}}_1 \rangle \cdot \langle \hat{\vec{\mu}}_2 \rangle - 3 (\langle \hat{\vec{\mu}}_1 \rangle \cdot \hat{r}) (\langle \hat{\vec{\mu}}_2 \rangle \cdot \hat{r})}{r^3}
\end{equation}
where 

\begin{equation}
    \langle \hat{\vec{\mu}}_1 \rangle (t) = \langle \hat{\vec{\mu}}_2 \rangle (t) = \mu \cos (\theta_0 - \theta_B) \hat{e} (t) + \mu \sin (\theta_0 - \theta_B) \left[ \cos \omega^\prime t \ \hat{\theta} (t) + \sin \omega^\prime t \ \hat{\varphi} (t) \right],
\end{equation}
as shown in eq. (\ref{eq:dipole_moment_dyn_gen_lowest_state}). Note that $\mu = g_J \mu_B (J-n)$ is the magnitude of the dipole moment. 

As discussed in the main text, when the timescale of the DDI strengths and other energy scales of the system are much longer than the timescales set by $\omega^\prime$ and $\Omega$, only the time-averaged DDI is significant. As a result, the atoms effectively only experience a time-averaged DDI over some time, $T \gg 2\pi/\Omega, \ 2\pi/\omega^\prime$, given by $\overline{V}_{dd} (\vec{r}) = (1/T) \int\limits_{0}^{T} dt V_{dd} (\vec{r},t)$. 

Now, $\langle \hat{\vec{\mu}}_1 \rangle (t) \cdot \langle \hat{\vec{\mu}}_2 \rangle (t) = \mu^2$, which is time-independent, so we only need to calculate the second term. Expanding the second term, we get: 

\begin{equation}\label{eq:ddi_second_term_expanded}
    \begin{split}
        &\langle \vec\mu_1 \cdot \hat{r}\rangle \langle \vec\mu_2 \cdot \hat{r}\rangle = \mu^2 \cos^2 (\theta_0 - \theta_B) (\hat{e} \cdot \hat{r})^2 + \dfrac{\mu^2 \sin^2 (\theta_0 - \theta_B)}{2} \left[ (\hat{\theta} \cdot \hat{r})^2 + (\hat{\varphi} \cdot \hat{r})^2 \right] \\
        &\hspace{0.5cm} + \dfrac{\mu^2 \sin^2 (\theta_0 - \theta_B) \cos 2\omega^\prime t}{2} \left[ (\hat{\theta} \cdot \hat{r})^2 - (\hat{\varphi} \cdot \hat{r})^2 \right] + \mu^2 \sin^2 (\theta_0 - \theta_B) \sin 2\omega^\prime t (\hat{\theta}  \cdot \hat{r}) (\hat{\varphi} \cdot \hat{r}) \\
        &\hspace{0.5cm} + 2\mu^2 \sin (\theta_0 - \theta_B) \cos (\theta_0 - \theta_B) \left[ \cos \omega^\prime t \left(\hat{e} \cdot \hat{r} \right) \left(\hat{\theta} \cdot \hat{r} \right) + \sin \omega^\prime t \left(\hat{e} \cdot \hat{r} \right) \left(\hat{\varphi} \cdot \hat{r} \right) \right]
    \end{split}
\end{equation}
For $T \gg 2\pi/\omega^\prime, \ 2\pi/\Omega$ and for $\Omega \neq \omega^\prime, \ 2\omega^\prime \text{ and } \omega^\prime/2$, we may use the orthogonality of trigonometric functions with different frequencies with respect to integration over $T$ to simplify the long-time average of equation (\ref{eq:ddi_second_term_expanded}). This ultimately gives us equation (\ref{eq:tilman-pfau-generalization-genOmega}) in the main text. 

\bibliographystyle{iopart-num}

\bibliography{References}
\end{document}